\documentclass[10pt,preprint,longnamesfirst]{aastex}

\slugcomment{ApJ, in press: 25 September 2002}

\shorttitle{Our Sun.\ V.\ A Bright Young Sun}
\shortauthors{Sackmann \& Boothroyd}

\newcommand{\iso}[2]{\hbox{${}^{#1}$#2}}
\newcommand{\gammaone}{\Gamma_{\!1}}

\begin{document}

\title{Our Sun.\ V. A Bright Young Sun Consistent with Helioseismology
 and Warm Temperatures on Ancient Earth and Mars}

\author{I.-Juliana Sackmann\altaffilmark{1} and
 Arnold I. Boothroyd\altaffilmark{2}}
\affil{W. K. Kellogg Radiation Laboratory 106-38,
 California Institute of Technology,\\
 Pasadena, CA 91125}
\email{ijs@caltech.edu}
\email{boothroy@cita.utoronto.ca}
\altaffiltext{1}{Present address: West Bridge Laboratory 103-33,
 California Institute of Technology, Pasadena, CA 91125}
\altaffiltext{2}{Present address: CITA, U. of Toronto, 60 St.\ George Street,
 Toronto, Ontario, Canada M5S 3H8}

\begin{abstract}

The relatively warm temperatures required on early Earth and Mars
have been difficult to account for via warming from greenhouse gases.
We tested whether this problem can be resolved for both Earth and Mars by
a young Sun that is brighter than predicted by the standard solar model (SSM).
We computed high-precision solar evolutionary
models with slightly increased initial masses of
$M_i = 1.01$ to~$1.07\;M_\odot$;
for each mass, we considered three different mass loss scenarios.
We then tested whether these models were consistent with
the current high-precision helioseismic observations.
The relatively modest mass loss rates in these models are consistent with
observational limits from young stars and
estimates of the past solar wind obtained from lunar rocks,
and do not significantly affect the solar lithium depletion.
For appropriate initial masses,
all three mass loss scenarios are capable of yielding a solar
flux 3.8~Gyr ago high enough to be consistent with water on ancient Mars.
The higher flux at the planets is due partly to the fact that a
more massive young Sun would be intrinsically more luminous, and partly
to the fact that the planets would be closer to the more massive young Sun.
At birth on the main sequence, our preferred initial mass
$M_i = 1.07\;M_\odot$ would produce a solar flux at the
planets 50\% higher than that of the SSM, namely, a flux
5\% {\it higher\/} than the present value
(rather than 30\% lower, which the SSM predicts).
At first (for $1 - 2$~Gyr), the solar flux would {\it decrease\/};
subsequently, it would behave more like the flux in the SSM, increasing
until the present.
%
We find that
all of our mass-losing solar models are consistent with
the helioseismic observations; in fact, our preferred mass-losing case
with~$M_i = 1.07\;M_\odot$
is in marginally (though insignificantly) better agreement with the
helioseismology than is the SSM\hbox{}.
The early solar mass loss of a few percent does indeed
leave a small fingerprint on the Sun's internal structure.
However, for helioseismology to significantly
constrain early solar mass loss would require higher accuracy in
the observed solar parameters and input physics, namely,
by a factor of~$\sim 3$ for the observed solar surface composition,
and a factor of~$\sim 2$ for the solar interior
opacities, the $pp$~nuclear reaction rate, and the diffusion constants for
gravitational settling.

\end{abstract}



\section{Introduction} \label{sec:intro}

Observations indicate that the Earth was at least
warm enough for liquid water to exist
as far back as 4~Gyr ago,
namely, as early as half a billion years after the formation of the Earth
\citep{CogH84,Moj+96,EilMA97,Erik82,Bow89,Nut+84};
in fact, there is evidence suggesting that Earth may have been
even warmer then than it is now
\citep{Kast89,OberMA93,Woese87,OhmF87,KnauE76,KarE86}.
These relatively warm temperatures required on early Earth are in apparent
contradiction to the dimness of the early Sun predicted by the
standard solar models.  This problem has generally been explained
by assuming that Earth's early atmosphere contained huge amounts of
carbon dioxide (CO$_2$), resulting in a large enough greenhouse effect to
counteract the effect of a dimmer Sun.  However, the recent work of
\citet{RyeKH95}
places an upper limit of 0.04~bar on the partial pressure of CO$_2$ in the
period from 2.75 to 2.2~Gyr ago, based on the absence of siderite in
paleosols; this casts doubt on the viability of a strong CO$_2$ greenhouse
effect on early Earth.
The existence of liquid water on early Mars has been even more of
a puzzle; even the maximum possible CO$_2$ greenhouse effect cannot
yield warm enough Martian surface temperatures
\citep{Kast91,KastWR93}.
These problems can simultaneously be resolved, for both Earth and Mars,
if the early Sun were brighter than predicted by the standard solar models.
This could be accomplished if the early Sun were slightly more massive
than it is now.

Helioseismic observations provide revolutionary precision for probing the
solar interior.  Helioseismic frequencies are observed with an accuracy
of a few parts in~$10^5$, allowing measurement of the sound speed profile
throughout most of the Sun's interior to an accuracy of a few parts
in~$10^4$
\citep{BasuPB00}.
This high precision permits one to search for subtle effects in the
interior structure of the present Sun resulting from
events in the distant past.  In particular, modest mass loss
(between 1\% and~7\% of the Sun's mass) early on the main
sequence might have left enough of a fingerprint on the
interior structure of the present Sun
to be detectable by helioseismological observations.

\subsection{Limits On Early Solar Mass Loss} \label{ssec:mdotlimit}

\subsubsection{Theoretical Solar Models and Helioseismology}
  \label{sssec:mdotmodelsun}

\citet{WilBS87}
first presented the hypothesis that stars like the Sun might
lose significant amounts of mass on the early main sequence.
\citet{GuzWB87}
were the first to compute solar models with such early
main sequence mass loss, namely, an extreme case with an initial mass
of~$2\;M_\odot$.  Such extreme mass loss (of~$\Delta M = 1\;M_\odot$)
turns out to be unrealistic, as discussed below; but small mass loss
cases cannot be ruled out at the present.
\citet{BSF91}
considered an initial solar mass of~$1.1\;M_\odot$, showing that this is
the upper limit allowed by the observed solar lithium depletion.
Some work has been carried out recently attempting to use helioseismology
to constrain early solar mass loss
\citep{GuzC95,MorPB97};
however, the first of these used rather crude solar interior models, and
the second yielded ambiguous results, as we discuss in detail below.

\citet{GuzC95}
were the first to attempt to use helioseismic observations to constrain
early solar mass loss; they
considered initial solar masses of 1.1 and~$2\;M_\odot$, concluding
that the $2\;M_\odot$ could be ruled out by helioseismic observations of
low-degree modes.  They also claimed that a mass loss timescale of 0.2~Gyr
was favored over a 0.45~Gyr timescale for the $1.1\;M_\odot$ case, based on
``frequencies for modes that probe just below the convection zone bottom
($l = 5 - 20$, $\nu = 2500 - 3500 \; \mu$Hz),''  but this latter conclusion
is rendered dubious by the fact that this is the particular region of the
Sun where one should expect the {\it worst\/} agreement in models
that do not attempt to include rotation-induced mixing
\citep[see, e.g.,][]{Rich+96,BrunTZ99,BasuPB00,BahPB01,TurckC+01b}.
Furthermore, these models of
\citet{GuzC95}
were based on relatively crude approximations to the physical processes
in the solar interior, particularly for the opacities.  They state
that their solar interior opacities used the
\citet{Iben75}
analytical fits to the opacity tables of
\citet{CoxS70},
with the electron scattering (esk) term in the Iben fit being multiplied
either by a factor of~1.5 to approximate the LAOL tables, or by a factor
of~2 to approximate the (early) OPAL opacities of
\citet{RogI92};
in addition, they adjusted the opacities {\it separately\/}
for each of their solar models, by adjusting the
$A_z$ term in the Iben fit (which mainly affects opacities at
$2 - 5 \times 10^6 \;$K) in order to yield a position for the base of
solar envelope convection at $R_{ce} \approx 0.711 - 0.712 \; R_\odot$
\citep{GuzC95}.
Even if they had in fact used the early OPAL opacity tables of
\citet{RogI92},
the models of
\citet{MorPB97}
suggest that this would be expected to yield solar models with
helioseismic disagreements 2~or 3 times worse than can be achieved
by models using the more recent OPAL opacities
\citep{IglR96,RogSI96};
we have shown that using the LAOL opacity tables
would result in even larger errors
\citep{BS02}.
\citet{GuzC95}
used the recent MHD equation of state tables
\citep{Dap+88},
albeit for a fixed value of $Z_{eos} = 0.02$
(the metallicity~$Z$ actually varied from $\sim 0.018$
to $\sim 0.021$ from the envelope to the center of their models), but the
equation of state is not very sensitive to metallicity, and we estimate that
this fixed-$Z$ approximation introduced errors in their thermodynamic
quantities of no more than a few parts in~$10^3$.  They considered
diffusion of hydrogen, helium, and four of the CNO isotopes, using
diffusion constants from
\citet{CoxGK89};
as they point out, the fact that this treatment probably overestimates
the extent of gravitational settling of heavy elements
\citep{ProfM91}
may compensate for the fact that they considered diffusion for
elements comprising only 62\% of the metallicity~$Z$.  Finally,
\citet{GuzC95}
compared models which had not been converged to the same final surface
$Z/X$ value; their $Z/X$ differences of up to~4\% would by themselves
yield significant (though not major) differences between the models,
largely through the effect on the opacities
\citep{BS02}.

\citet{MorPB97}
also used helioseismology to test an initial solar mass of~$1.1\;M_\odot$,
comparing sound speed and density profiles in their solar models to
profiles obtained via an inversion from helioseismic frequencies.
Their solar models incorporated much more up-to-date input physics
(though they also used the fixed-$Z$ approximation in the equation of
state, with $Z_{eos} = 0.019$),
and they tested some different formulations of the input physics;
but their tests of the effects of early solar mass loss yielded results
that were ambiguous at best.  They first considered models using the
CEFF equation of state
\citep{ChrD92},
diffusion constants from
\citet{MicP93},
and the more recent OPAL opacities
\citep{IglR96,RogSI96};
however, in these models
\citet{MorPB97}
also used the fixed-$Z$ approximation when interpolating the
{\it opacities}, which can lead to significant opacity errors
and thus significant effects on the solar models, as they themselves showed
\citep[see also][]{BS02}.
\citet{MorPB97}
compared mass-losing models with an initial mass of
$M_i = 1.1 \; M_\odot$ to a standard solar model ($M_i = 1.0 \; M_\odot$),
testing both a short (0.2~Gyr) and a long (0.45~Gyr)
mass loss timescale.  They
found that the long mass loss timescale yielded fractional shifts of up
to~$\sim 0.001$ in the sound speed profile relative to the standard
solar model (a significant but not major effect), while the short mass
loss timescale had an insignificant
effect (of~$\sim 0.0002$, roughly one fifth as large).
They next considered models with the OPAL equation of state
\citep{RogSI96},
and where the variation of metallicity~$Z$ inside the star {\it was\/}
considered when interpolating in the OPAL opacities (each of these latter
two changes yielded shifts of up to 0.001 in the sound speed profiles of
their standard solar models).  They compared a model with an initial mass
$M_i = 1.1 \; M_\odot$ and their short (0.2~Gyr) mass loss timescale to a
standard solar model, finding differences of up to~$\sim 0.001$ in the
sound speed profile between them --- 5~times as large as the effect from the
same comparison with the first set of models (and large enough to suggest
the possibility of highly significant effects from a longer mass loss
timescale).  Such large disagreements between their two sets of
results for the differential effects of early solar mass
loss suggest that their mass-losing models may have had much larger
numerical or convergence errors than they estimated for their standard
solar models; and indeed their second short-timescale mass-losing model
shows sound speed disagreements near the surface that might be interpreted
as being due to poor convergence to the present solar luminosity, radius,
and/or surface~$Z/X$ --- though even very poor convergence would not
expected to yield such large interior disagreements
\citep{BS02}.

The goal of the present work was to resolve the above ambiguity as to
what constraints might be placed on early solar mass loss by helioseismic
observations, by computing models with both up-to-date input physics
and high numerical precision.  We considered additional observational
constraints on solar mass loss (as discussed in \S~\ref{sssec:mdotobs}
below), and tested a number of different initial solar
masses and mass loss timescales.  We also compared the effects of mass
loss to the effects of other reasonable variations in the solar input
parameters and input physics --- these latter effects were discussed in
detail in our companion paper ``Our Sun~IV''
\citep{BS02}.
Finally, we present the effects of the various mass loss cases on the solar
flux reaching Earth and Mars as a function of time.

\subsubsection{Other Observational Constraints}
  \label{sssec:mdotobs}

Presently, the Sun is experiencing only a negligible amount of mass loss:
the solar wind removes mass at a rate
$\sim 3 \times 10^{-14}\; M_\odot$~yr$^{-1}$.  If this mass loss rate
had been constant over the last 4.5~Gyr, the young Sun would have
been more massive by only $\sim 10^{-4}\; M_\odot$.  The contemporary solar
wind has been observed only for a few decades, and has been found to be
a highly variable phenomenon --- all properties, including flux, velocity,
and composition vary significantly
\citep{GeiB91}.
The lunar surface material carries the signature of the solar wind
irradiation over the past several Gyr;
measurements of noble gas isotopes in lunar samples suggest that the
average solar wind flux over the past $\sim 3$~Gyr was an order
of magnitude higher than it is today
\citep{Gei73,GeiB91,Ker+91}.
This implies a total solar mass loss of $\sim 10^{-3}\; M_\odot$ over
the past 3 to 4~Gyr (the age of the oldest available lunar material).
Some older, solar-flare irradiated grains from meteorites imply an
early solar flare activity about $10^3$~times that of the present Sun
\citep{CafHS87};
the associated solar wind may have been enhanced by a similar factor
of~$\sim 10^3$, most likely during the first $\sim 1$~Gyr of the Sun's life
on the main sequence
\citep{Whit+95},
implying a total mass loss during this first 1~Gyr period of as much
as $\sim 0.03\; M_\odot$ (if the average mass loss rate throughout
that period was indeed $\sim 10^3$ times the present rate of
$3 \times 10^{-14}\; M_\odot$yr$^{-1}$).  Such a change in the solar
mass would be sufficient to cause a significant increase in the luminosity
of the young Sun.

Since the Sun is a typical main sequence star, it is reasonable to assume
that mass loss rates in the young Sun would be similar to those in other
young solar-type main sequence stars.
There have been several attempts to measure mass loss in early main
sequence stars.  It is observationally a very challenging task.
\citet{Brown+90}
attempted to obtain mass loss rates for 17 young main sequence stars
somewhat hotter and more massive than the Sun (A~and~F dwarfs), finding
upper limits to the mass loss rates of $10^{-10}$ to $10^{-9}\; M_\odot$/yr;
these limits are even less constraining than the highest solar mass loss
rate suggested by the meteoritic and lunar data.
\citet{GaiGB00}
used 3.6~cm VLA observations to place more stringent upper limits of
$\dot M \lesssim 5 \times 10^{-11} \; M_\odot \,$yr$^{-1}$
on mass loss rates of three young main sequence stars of roughly solar
mass ($\pi^{01}$~UMa, $\kappa^1$~Cet, and $\beta$~Com) --- as discussed
in \S~\ref{ssec:mdotused}, the largest initial solar mass that we
consider, namely $1.07\;M_\odot$,
would require early solar mass loss rates that are marginally
consistent with these more stringent limits.

\citet{Wood+01}
recently obtained HST observations of \ion{H}{1} Ly$\alpha$ absorption from
the region where the stellar wind collides with the interstellar medium,
using these to measure the stellar wind from the sun-like star $\alpha$~Cen.
They found a mass loss rate roughly twice as large as that of the Sun;
note that $\alpha$~Cen is slightly {\it older\/} than the Sun (with an
age of $\sim 5$~Gyr), as well as being very slightly more massive
($M \approx 1.08 \; M_\odot$).
They also found an upper limit roughly ten times lower for its cooler,
less massive companion Proxima Cen.  A similar method had earlier
been used by
\citet{WoodL98}
to look at four other main sequence stars cooler and less
massive than the Sun (finding stellar winds of roughly the same order
of magnitude as the solar wind).
Such a method applied to young, Sun-like stars holds promise
for placing stringent limits on early main sequence mass loss.

Very recently,
\citet{Wood+02}
demonstrated that mass loss rates in GK~dwarfs (measured using the above
method) are generally correlated with their X-ray flux
(measured by ROSAT, and using the HIPPARCOS distances).  For 7 of the
9 stars they looked at, there was a fairly tight correlation, with
$\dot M \propto F_{\!X}^{1.15\pm0.20}$ over a range of 2~orders of
magnitude in $\dot M$ and~$F_{\!X}$.  They then used the estimates from
\citet{Ayres97}
of the correlation of X-ray flux with rotational velocity
$F_{\!X}\propto V_{rot}^{2.9\pm0.3}$
and of rotational velocity with age
$V_{rot} \propto t^{-0.6\pm0.1}$
to obtain the first empirically derived relation describing the mass-loss
evolution of cool main-sequence stars like the Sun, namely,
$\dot M \propto t^{-2.00\pm0.52}$.
They point out that the above correlations do not really apply to very
young stars (age${} \lesssim 0.3$~Gyr), and assume instead that the
maximum mass loss rate at early times corresponds to the maximum X-ray
flux observed in Sun-like stars, namely, $10^3$~times the solar value.
Given the Sun's present mass loss rate of
$\dot M \sim 3 \times 10^{-14} \; M_\odot \,$yr$^{-1}$,
this would imply total solar mass loss of order~$0.01\;M_\odot$
(albeit with an uncertainty of a factor of~5 or so), with
most of this mass loss taking place in the first fraction of a Gyr
of the Sun's lifetime.  The effects of such a mass loss law will be
investigated in more detail in a future paper
\citetext{A.~I.~Boothroyd \& \hbox{I.-J.}~Sackmann, in preparation},
but we make some rough estimates of the probable effects in
\S~\ref{sec:results} and~\S~\ref{sec:concl}.

The observed depletion of lithium in the Sun provides a stringent upper limit
to the total solar mass loss of $\Delta M \sim 0.1\; M_\odot$;
i.e., the initial solar
mass~$M_i$ (4.5~Gyr ago) is constrained to be $M_i \lesssim 1.1\; M_\odot$
\citep{BSF91}.
However, this is much too generous an upper limit.  There are additional
mechanisms that can deplete solar lithium.  One mechanism, namely,
pre-main-sequence lithium depletion (during the Sun's initial contraction
phase), was taken into account in our mass-losing solar models
(for the standard Sun, this depletion was a factor of~$\sim 20$,
as discussed below and in our companion paper ``Our Sun~IV''
[\citealt{BS02}]).
Another mechanism is rotation-induced turbulent mixing,
which probably is the major cause of the main-sequence
lithium depletion; however, rotation models have free parameters, and
can be fitted to {\it any\/}
required amount of solar lithium depletion
\citep[see, e.g.,][]{Schat77,LebM87,Pins+89,CharVZ92,Rich+96}.
Also, it has been shown that mass loss cannot be the major
contributor to the observed lithium depletions in the young Hyades cluster
\citep{SwenF92};
this constraint is discussed in more detail in \S~\ref{sssec:hyrest} below.

An even more stringent upper limit to the Sun's initial mass is imposed
by the requirement that the early Earth not lose its water via a
{\it moist greenhouse effect}, which would occur if the solar flux at Earth
were more than 10\% higher than its present value
\citep{Kast88}
--- a moist greenhouse occurs when the stratosphere becomes wet, and
H$_2$O is lost through UV dissociation and the subsequent loss of
hydrogen to space.  This solar flux limit corresponds to an upper
limit on the Sun's initial mass of $M_i \lesssim 1.07\; M_\odot$,
which is the most stringent upper limit on the Sun's initial mass.

The only strong lower limit on $M_i$ comes from the fact that the Sun is
converting matter into energy and radiating it away;
$\Delta E = L \Delta t = \Delta M c^2$, where $\Delta E$ is the total
energy radiated away, $L$~is the average solar luminosity
(including the neutrino luminosity)
$\Delta t$ is the $\sim 4.5$~Gyr duration of the nuclear burning,
$\Delta M$ is the amount of mass converted into energy, and $c$~is
the speed of light (note that elsewhere in the paper we use ``$c$''
to denote the adiabatic sound speed).  At present, mass
is radiated away as photons and
neutrinos at a rate slightly over twice the solar wind mass loss rate.
For the standard
solar model, the Sun's average luminosity over the last 4.5~Gyr was
about 0.85 times its present luminosity.  It follows that
$\Delta M \approx 3 \times 10^{-4}\; M_\odot$ from radiation losses
alone (i.e., that $M_i \gtrsim 1.0003\; M_\odot$).  Such a minor amount of
mass loss has a negligible effect on the early solar luminosity.

There are also considerations that put soft lower limits on the Sun's
initial mass~$M_i$.  If the present observed solar wind rate of
$\sim 3 \times 10^{-14}\; M_\odot$/yr had been constant over the Sun's
history, the total amount of solar wind mass loss would have
been only $\sim 1.4 \times 10^{-4}\; M_\odot$; including the
$\Delta M \approx 3 \times 10^{-4}\; M_\odot$ from radiation losses
would imply $M_i \sim 1.0004\; M_\odot$.  However,
measurements of the noble gases implanted in lunar samples suggest an
average solar wind flux over the past $\sim 3$~Gyr an order of magnitude
higher than at present,
\citep{Gei73,GeiB91,Ker+91},
implying a total solar mass loss over that period of~$\sim 0.001\; M_\odot$,
i.e., a solar mass 3~Gyr ago of~$M({\rm -3\>Gyr}) \sim 1.001\; M_\odot$ ---
note that $M(-t)$ is used to refer to the solar mass at $t$~years before
the present.
The $\sim 3$~Gyr age of these lunar rocks means that they place no limits
on {\it earlier\/} solar mass loss, so that all one can say is
that $M_i \ge M({\rm -3\>Gyr})$.
Older, solar-flare irradiated grains from meteorites imply
early solar flare activity about $10^3$~times that of the present Sun
\citep{CafHS87},
which might possibly correspond to similarly high mass loss rates
during the first $\sim 1$~Gyr period of the Sun's life,
but cannot be used to provide any sort of limit.  The
semi-empirical mass loss formula recently presented by
\citet{Wood+02}
suggests total solar mass loss of order~$0.01\;M_\odot$, but with an
uncertainty of about a factor of~5, as discussed above.

Another limit on the Sun's initial mass
comes from the requirement that Mars was warm enough
for liquid water to exist 3.8~Gyr ago (at the end of the late heavy
bombardment period).  According to
\citet{Kast91}
and
\citet{KastWR93},
this requires a solar flux (at Mars) 3.8~Gyr ago at least 13\% larger
than that from the standard solar model, in order to make it possible
for a CO$_2$ greenhouse effect on Mars
to be able raise the temperature to~$0^\circ\,$C\hbox{}.
Such an increase in flux would correspond to a mass of the Sun at that time
of $M({\rm -3.8\>Gyr}) \gtrsim 1.018\; M_\odot$.
Since the lunar rock measurements
constrain the Sun's mass $\sim 3$~Gyr ago to
be $M({\rm -3\>Gyr}) \sim 1.001\; M_\odot$,
the Sun's average mass loss rate between 3.8 and 3~Gyr ago would be
$\dot M \gtrsim 2 \times 10^{-11}\; M_\odot$/yr.  If this same mass loss rate
also occurred throughout the period from the Sun's birth $\sim 4.6$~Gyr ago
until 3.8~Gyr ago, this would imply an initial solar mass
of $M_i \gtrsim 1.033\; M_\odot$.  Note that this lower limit assumes that
the only greenhouse effect on early Mars is due to~CO$_2$.
If a smog-shielded ammonia greenhouse could exist on early Mars,
such as that proposed for the early Earth by
\citet{SagC97},
then this lower limit on~$M_i$ might be softened or eliminated.

\subsubsection{The Swenson-Faulkner Hyades Mass Loss Constraint}
  \label{sssec:hyrest}

\citet{SwenF92}
established that mass loss could not be
the {\it major\/} cause of the main-sequence lithium depletion
for stars in the Hyades cluster.
Their result has frequently been mis-quoted and misunderstood;
it has often been used to rule out the possibility of
mass loss during the
Sun's early main sequence phase.  However, 
their result does {\it not\/} rule out relatively small
amounts of mass loss for either the Hyades or the Sun.

For the Hyades cluster, which is 0.6~Gyr old,
lithium abundances in many stars have been observed,
exhibiting a fairly tight relationship between a star's lithium abundance
and its surface temperature --- the observed lithium abundance drops off
steeply with decreasing surface temperature, below $\sim 6000$~K\hbox{}.
\citet{SwenF92}
considered lithium depletion due both to pre-main-sequence burning and to
main sequence mass loss.  
They found that the observed lithium-temperature relationship
could not be accounted for by pre-main-sequence
lithium depletion alone, but that it could be accounted for fairly well
if one added main sequence mass loss.  However, they found that
all the stars with surface temperatures below 5500~K would then have to
have nearly identical initial masses (with a wide range of mass loss
rates).  Such a distribution of initial stellar masses, with a high,
narrow peak in the distribution near $1.1\; M_\odot$, is unrealistic.
This argument has been widely misquoted, to rule out early main
sequence mass loss in stars (including the Sun).

The Swenson-Faulkner conclusion applies {\it only\/} if one is trying
to match the Hyades lithium depletions {\it without including
rotation-induced mixing}.  As soon as one includes the latter as a
major component, one can reproduce the observed lithium-temperature
relation of Hyades stars by choosing suitable values for the adjustable
parameters in the rotational mixing formalism
\citep[see, e.g.,][]{Schat77,LebM87,Pins+89,CharVZ92,Rich+96}.
Stellar rotation is ubiquitous in young stars, and is commonly assumed
to be the cause of all main sequence lithium depletion; the
presence of a relatively
small amount of mass loss merely requires that the large
lithium depletion due to rotation be decreased by a small amount
(by small changes in the adjustable parameters for rotational mixing).
For the Hyades, even a mass loss as large as
$\Delta M = 0.07\; M_\odot$ in a star near $1\; M_\odot$
would imply a lithium depletion factor due to mass loss alone
of only~$\sim 5$,
\citep[according to the models of][]{SwenF92},
and would still require a depletion factor due to rotational mixing
of~$\sim 15$ in order to reproduce the observed
lithium-temperature relation.
For the Sun, pre-main sequence burning yields a lithium depletion factor
of order~20
\citep{BS02};
as discussed in \S~\ref{sssec:lithium}, combining pre-main sequence lithium
burning with an early solar mass loss of $\Delta M = 0.07\; M_\odot$
increases this by less than a factor of~2, i.e., a combined lithium depletion
factor of~$\sim 30 - 40$, still much smaller than the total observed lithium
depletion of~$160 \pm 40$
\citep{GreS98}.
Rotation would be responsible for the
remaining lithium depletion.  Mass loss of order $\Delta M = 0.07\; M_\odot$
or less {\it is consistent\/} with the Hyades lithium observations, i.e.,
does not require an unrealistic initial stellar mass distribution.

In their later work,
\citet{Swen+94}
found they were able to reproduce the Hyades lithium-temperature relation
by pre-main-sequence depletion alone, provided that the oxygen abundance
was assumed to be at the upper limit of the observed range (and using
the most up-to-date OPAL and Alexander opacities).  However, other
clusters such as NGC~752, M67, or NGC~188 {\it cannot\/} be explained
via pre-main-sequence lithium
depletion alone --- their observed lithium depletions are much {\it larger\/}
than those of the Hyades
\citep[see, e.g.,][]{HobP88,Bal95},
while their pre-main-sequence depletions
would be significantly {\it smaller\/} (due to their lower metallicities).
The same is true of the Sun.  In other words, the observed lithium
depletions demand main sequence depletion, which would be largely due
to rotation-induced mixing, with possibly a small effect from mass loss
as well.


\section{Methods} \label{sec:methods}

\subsection{Mass Loss of the Young Sun} \label{ssec:mdotused}



\begin{figure}[t]
 \epsscale{0.67}
 \plotone{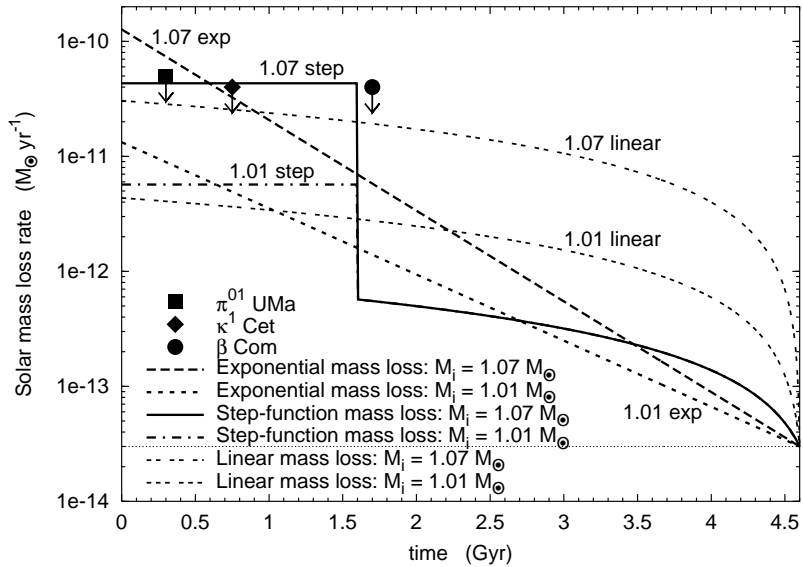}

\caption{Mass loss rates as a function of time for our ``exponential,''
``step-function,'' and ``linear'' solar mass loss cases.  The lowest and
the highest mass loss cases that we considered are shown ($M_i = 1.01$ and 
$1.07\;M_\odot$, respectively).  Mass loss upper limits for the young
Sun-like stars $\pi^{01}$~UMa, $\kappa^1$~Cet, and $\beta$~Com are from
\citet{GaiGB00}.
}

 \label{fig:mdot}

\end{figure}


We computed mass-losing solar models having initial
masses~$M_i = 1.01$,
1.02, 1.03, 1.04, 1.05, 1.06, and~$1.07\;M_\odot$; as discussed
in~\S~\ref{sssec:mdotobs},
an initial mass of~$1.07\;M_\odot$ is
the upper limit consistent with the requirement that the young Earth not
lose its water via a moist greenhouse effect
\citep{Kast88}.
We considered three different forms for early solar mass loss, which
we call ``exponential,'' ``step-function,'' and ``linear''; for each
of these, two limiting cases are displayed in Figure~\ref{fig:mdot}
(i.e., cases with initial solar masses of 1.01 and $1.07 \; M_\odot$).

In the ``exponential'' mass loss case, the mass loss rate starts out
high and declines exponentially, with an initial mass loss
rate~$\dot M_0$ and decay time constant~$\tau$ chosen
such as to give the present observed solar mass loss rate of
$\dot M = 3 \times 10^{-14} \; M_\odot\,$yr$^{-1}$
at the Sun's present age.  In other words,
$\dot M(t) = \dot M_i \, e^{-t/\tau}$, with
$\dot M_i = 1.33 \times 10^{-11}$ to 
$1.27 \times 10^{-10} \; M_\odot\,$yr$^{-1}$ and
$\tau = 0.755$ to 0.551~Gyr, for initial
solar masses of 1.01 to $1.07 \; M_\odot$, respectively.
This exponential mass loss case
is our most conservative
one: a simple mass loss scenario that is still reasonably
consistent with the observed lunar mass loss constraint.
This scenario yields average mass loss rates over the past 3~Gyr of 13 to~42
times the present value, for initial solar masses of 1.01 to 
$1.07 \; M_\odot$, respectively.  These average mass loss
values are reasonably consistent with measurements of noble gases in
lunar rocks, which suggest a mass loss rate an order-of-magnitude
higher than the present value.
Since Figure~\ref{fig:mdot} is a log-linear plot, these ``exponential''
mass loss cases appear as straight lines.

The ``step-function'' mass loss rate was chosen to have a constant
high value during the period before the lunar rock observations apply,
namely, the first 1.6~Gyr of the Sun's life; over the remaining
3~Gyr of the Sun's life (up to the present), a mass loss rate averaging
ten times the present value was assumed,
declining linearly over this period to reach the present solar mass
loss rate at the present solar age.  This is the most extreme case
which is still consistent with the observed lunar rock mass loss
constraints: it
keeps the solar flux as high as possible for as long as possible.
For the first 1.6~Gyr, this scenario has constant mass loss rates of
$\dot M = 5.69 \times 10^{-12}$ to 
$4.32 \times 10^{-11} \; M_\odot\,$yr$^{-1}$, for initial
solar masses of 1.01 to $1.07 \; M_\odot$, respectively.

In the ``linear'' mass loss case, the mass loss rate starts out high
and declines slowly and linearly, to reach the present solar mass
loss rate at the present solar age.
This was chosen as our most radical case, 
with maximum impact on the Sun's internal structure.  
Due to the linear decline, the mass loss rate remains high
throughout most of the Sun's lifetime, remaining of the same order
as the initial mass loss rate ($\dot M_i = 4.35 \times 10^{-12}$ to 
$3.04 \times 10^{-11} \; M_\odot\,$yr$^{-1}$, for
initial solar masses of 1.01 to $1.07 \; M_\odot$, respectively).
During the past 3~Gyr, the mass loss rate for this ``linear'' case
is much higher than for the other mass loss cases above, violating the
observed lunar mass loss constraints
(the ``linear'' case has average mass loss
rates over the past 3~Gyr of 50 to~330 times the present rate, for
initial solar masses of 1.01 to $1.07 \; M_\odot$, respectively).
The ``linear'' mass loss cases appear as curved lines in
Figure~\ref{fig:mdot}.

\subsection{Physical Inputs to our Solar Models} \label{ssec:inputs}

The solar evolution program is discussed in detail in our companion
paper ``Our Sun~IV''
\citep{BS02};
we provide only a brief summary here.  We used the OPAL equation of state
\citep{RogSI96},
extended to lower temperatures by the MHD equation of state
\citep{Dap+88}.
The 1995 OPAL opacities
\citep{IglR96}
were used for $\log\,T > 4$;
since these opacities (as well as both sets of equation of state tables)
were based on the heavy element composition of
\citet{GreN93},
this mixture was used in order to obtain self-consistent solar models
(along with their recommended value $Z/X = 0.0245$ in the present solar
envelope).  At lower temperatures ($\log\,T < 4$), the
\citet{AlexF94}
molecular opacities were used.  Both the equation of state and the opacities
were interpolated in metallicity as
well as in hydrogen abundance, temperature, and density, in order to
take into account metallicity variations due to diffusion and nuclear
burning.

For the ``exponential'' mass loss cases, we also computed an alternate
set of solar models based on the recent solar composition observations of
\citet{GreS98},
which yield $Z/X = 0.023$ in the present solar envelope; for
this composition, we obtained appropriate OPAL opacities via the
online opacity computation feature of the OPAL web page\footnote{
\url[http://www-phys.llnl.gov/Research/OPAL/]{http://www-phys.llnl.gov/Research/OPAL/}}.
Also, in these models we performed the switchover between the OPAL and MHD
equations of state at $\log\,T \sim 4$ (rather than at
$\log\,\rho \sim -2$, which corresponds to $\log\,T \sim 5.5$ in the Sun).
As discussed in
\citet{BS02},
the switchover region between the two equations of state was wide enough, and
the equations of state were similar enough in the chosen switchover regions,
that artifacts introduced by the switchover should not be significant
compared to the size of the inconsistencies in the OPAL equation of state.

We used the NACRE nuclear reaction rate compilation
\citep{Ang+99},
supplemented by the \iso7{Be} electron capture rates of
\citet{GruB97}.
Weak screening
\citep{Sal55}
was used --- note that it is a very good approximation to the exact
quantum mechanical solution for solar conditions
\citep[see, e.g.,][]{BahCK98,GruB98}.
All of the stable isotopes up to and including~\iso{18}O were considered
in detail, except for deuterium (which was assumed to have been burned
to~\iso3{He}).  The other isotopes up to~\iso{28}{Si} were included in
the code, but not in the nuclear reaction network,
since there are no significant effects under solar
conditions (except for~\iso{19}F, which was assumed to be in CNO-cycle
nuclear equilibrium for nuclear rate purposes).
Neutrino capture cross sections were taken from
\citet{BahU88},
except for the \iso8B-neutrino cross section for capture on \iso{37}{Cl},
where the more recent value (5\%~higher) of
\citet{Aufder+94}
was used.

A set of subroutines\footnote{
These subroutines are available from Bahcall's web page:
\url[http://www.sns.ias.edu/~jnb/]{http://www.sns.ias.edu/${}^{\sim}$jnb/}}
were kindly provided to us
\citetext{M.~H.~Pinsonneault 1999, private communication}
that take into account the diffusion (gravitational settling) of helium and
heavy elements relative to hydrogen
\citep[see also][]{ThoulBL94,BahPW95}.

A present solar mass of $M_\odot = 1.9891 \times 10^{33}$~g
\citep{CohT86}
was used, and a solar radius at the photosphere
($\tau = 2/3$) of $R_\odot = 695.98$~Mm
\citep{UlrR83,Gue+92}.
Our solar luminosity value of
$L_\odot = 3.854 \times 10^{33}$~erg~s$^{-1}$
\citep{SBK93}
is close (less than 1-$\sigma$) to the more recent value of
\citet{BahPB01};
as discussed in our companion ``Our Sun~V'' paper
\citep{BS02},
such a luminosity difference has negligible effect on the solar structure
(and only a minor effect on the neutrino rates).
We used a total solar age of $t_\odot = 4.6$~Gyr, measured from the Sun's
birth on the pre-main-sequence Hayashi track; this is only just outside the
range 4.55~Gyr${} < t_\odot < 4.59$~Gyr (in effect,
$t_\odot = 4.57 \pm 0.01$~Gyr) allowed by meteoritic ages
\citep{BahPW95},
sufficiently close that a more precise age value would have had very little
effect on the solar structure and helioseismology
\citep[see][]{BS02}.
Note that the earlier total solar (and solar system) age estimate
$\tau_{ss} = 4.53 \pm 0.03$~Gyr of
\citet{Gue89}
is consistent with the limits of
\citet{BahPW95}.
Our models took about 40~Myr to reach the zero age main sequence (ZAMS),
the point at which nuclear reactions in the core provide essentially all
the Sun's luminosity, and the pre-main-sequence contraction stops; this
pre-main-sequence timescale implies that the {\it total solar age\/}~$t_\odot$
used in this paper can be converted into a {\it main sequence\/} solar
lifetime by subtracting about 0.04~Gyr
\citep[this was also pointed out by][]{Gue89}.

We investigated the effects of using two different zonings.  Our
coarse-zoned solar models had about 2000 spatial zones in the
model, and about 200 time steps in the evolution from the zero-age
main sequence to the present solar age (plus about 800 time steps on
the pre-main-sequence), comparable to other work in this field
\citep[e.g., a factor of~2 more than][]{MorPB97};
mass-losing solar models might require several times
as many timesteps, due to the constraint that mass loss effects be kept
small over one timestep.  Typically, these models were converged to
match the solar luminosity, radius, and surface~$Z/X$ value
to within a few parts in~$10^5$;
a few cases where convergence was slow reached only about a part in~$10^4$.
Because the inconsistencies in the mass-losing solar models of
\citet{MorPB97}
had suggested the possibility that such models might tend to have numerical
inaccuracies larger than in standard solar models, as discussed in
\S~\ref{sssec:mdotmodelsun}, we also computed a number of
fine-zoned models, with $10\,000$ spatial zones and 1500
main-sequence time steps (plus 6000 pre-main-sequence time steps)
--- a factor of 5 increase in both spatial and temporal precision.
It was indeed more difficult to converge mass-losing solar models to
precise solar luminosity, radius, and surface~$Z/X$ values --- the
standard solar models were converged to these values with an accuracy
nearly a factor of ten better than the typical mass-losing models --- but
comparing the last few models in a convergence sequence revealed no
untoward effects.  
Considering the much larger amounts of CPU-time required, there seemed to be
no point in attempting to match the solar luminosity, radius, and surface
$Z/X$ values any better for fine-zoned cases than for coarse-zoned ones.
(A fine-zoned
converged mass-losing solar model took a few weeks of CPU-time on a
fairly high-performance ES40 computer, as compared to about half a day for
a coarse-zoned case, while a standard solar model without mass loss took only
about a fifth as long in each case; these times were roughly tripled on a
450~MHz Pentium~III PC\hbox{}.)
As discussed in our companion ``Our Sun~IV'' paper
\citep{BS02},
even the worst of the above convergence accuracies has a negligible effect
on the solar sound speed profile: up to 1 or 2 parts in~$10^4$ in the
convective envelope, and a few parts in~$10^5$ below it.
The fine-zoned mass-losing models differed from the corresponding
coarse-zoned ones by the same negligibly small systematic shift
(less than a part in~$10^4$ in the sound speed) that had been found in
standard solar models
\citep[see][]{BS02};
we concluded that our properly converged coarse-zoned mass-losing models are
perfectly adequate, and do not suffer from the problems encountered by the
mass-losing solar models of
\citet{MorPB97}.

Even the largest of the mass loss rates we considered (3~orders of magnitude
larger than the present solar wind) is still quite small in absolute terms
(for example, asymptotic giant branch stars encounter mass loss rates 6~orders
of magnitude larger still, and high-mass main sequence stars also have very
high mass loss rates).  Nonetheless, over a reasonable model timestep,
such mass loss would cause a mass layer near the solar surface to move
outwards significantly.  We therefore used an outer boundary condition for
our models at a point $\sim 10^{-3}\;M_\odot$ inwards from the solar surface,
computing sets of static envelopes for the region exterior to this point
(and requiring timesteps short enough that changes at this outer boundary
point remained small from one timestep to the next).  Since the mass loss
timescales for these outer layers are always at least 4~orders
of magnitude longer than their thermal timescales, this should be a
good approximation; and ignoring energy changes in this small outer region
should introduce fractional errors in the solar luminosity at the level of
$10^{-6}$ or less.  Test cases where this outer boundary point was
chosen differently showed no effect, as expected.

We compared our solar models to profiles of the solar sound
speed~$c_\odot$, density~$\rho_\odot$, and adiabatic index~$(\gammaone)_\odot$
obtained from the helioseismic reference model of
\citet{BasuPB00}\footnote{
 From the denser-grid machine-readable form of their Table~2, at
\url[http://www.sns.ias.edu/~jnb/]{http://www.sns.ias.edu/${}^{\sim}$jnb/}},
which they obtained by inversion from the helioseismic frequency
observations.  In the inversion process, a standard solar model is
required, but
\citet{BasuPB00}
demonstrated that the resulting $c_\odot$ and~$\rho_\odot$ profiles
of the helioseismic reference model are
relatively insensitive to uncertainties in the standard solar model
used for this purpose (except for uncertainties in~$R_\odot$).
They estimated a net uncertainty
of few parts in~$10^4$ for the sound speed~$c_\odot$ and adiabatic
index~$(\gammaone)_\odot$, and a few parts
in~$10^3$ for the density~$\rho_\odot$.  However, in the Sun's core
($r \lesssim 0.1\;R_\odot$), systematic uncertainties in the
helioseismic sound profile are increased by a factor of~$\sim 5$; this was
demonstrated by
\citet{BahPB01},
who compared helioseismic inversions of different helioseismic data sets.
We used their comparison to estimate the $r$-dependence of the systematic
error in~$c_\odot$ in the core and in the convective
envelope (namely, a fractional systematic
error decreasing linearly from~0.0013 at $r = 0.05\;R_\odot$
to 0.0003 at $r = 0.2\;R_\odot$, constant from there to $r = 0.72;R_\odot$,
then increasing linearly to~0.00052 at $r = 0.94;R_\odot$).
For~$c_\odot$, this systematic error can be significantly larger than the
statistical errors quoted in Table~2 of
\citet{BasuPB00},
and we combined the two in quadrature to get the total
fractional error~$( \sigma_c / c )$
for the purpose of calculating weighted rms differences --- the
rms fractional difference in~$c$ is given by
$\left( \left\{ \sum \left[ ( \Delta c / c )
   / ( \sigma_c / c ) \right]^2 \right\}
 / \left\{ \sum \left[ 1 / ( \sigma_c / c ) \right]^2 \right\} \right)^{1/2}$.
For $(\gammaone)_\odot$ and~$\rho_\odot$, the systematic errors are comparable
to or smaller than the statistical ones, and the statistical errors
sufficed for calculating weighted rms differences.

We present all our sound speed and density profiles in terms of
differences relative to the observed helioseismic reference profiles.
This choice of presentation not only allows one to see the effects of
the choice of initial mass and mass loss type, but also the extent to
which the models agree with the helioseismic observations.


\section{Results and Discussion} \label{sec:results}

For comparison with our solar mass loss rates, we used
the most recent observed upper limits on stellar mass loss rates from
three young Sun-like stars (namely,
$\dot M \lesssim 5 \times 10^{-11} \; M_\odot \,$yr$^{-1}$, from
$\pi^{01}$~UMa, $\kappa^1$~Cet, and $\beta$~Com), as presented by
\citet{GaiGB00}.
Even our highest mass loss
cases are very close to being consistent with these limits,
as is illustrated in Figure~\ref{fig:mdot}.

\subsection{Testing Mass Loss Models Via Helioseismology}
\label{ssec:testhelio}

\subsubsection{Sound Speed and Density Profiles}  \label{sssec:profiles}

\begin{figure}[tp]
  \epsscale{1.11}
 \plottwo{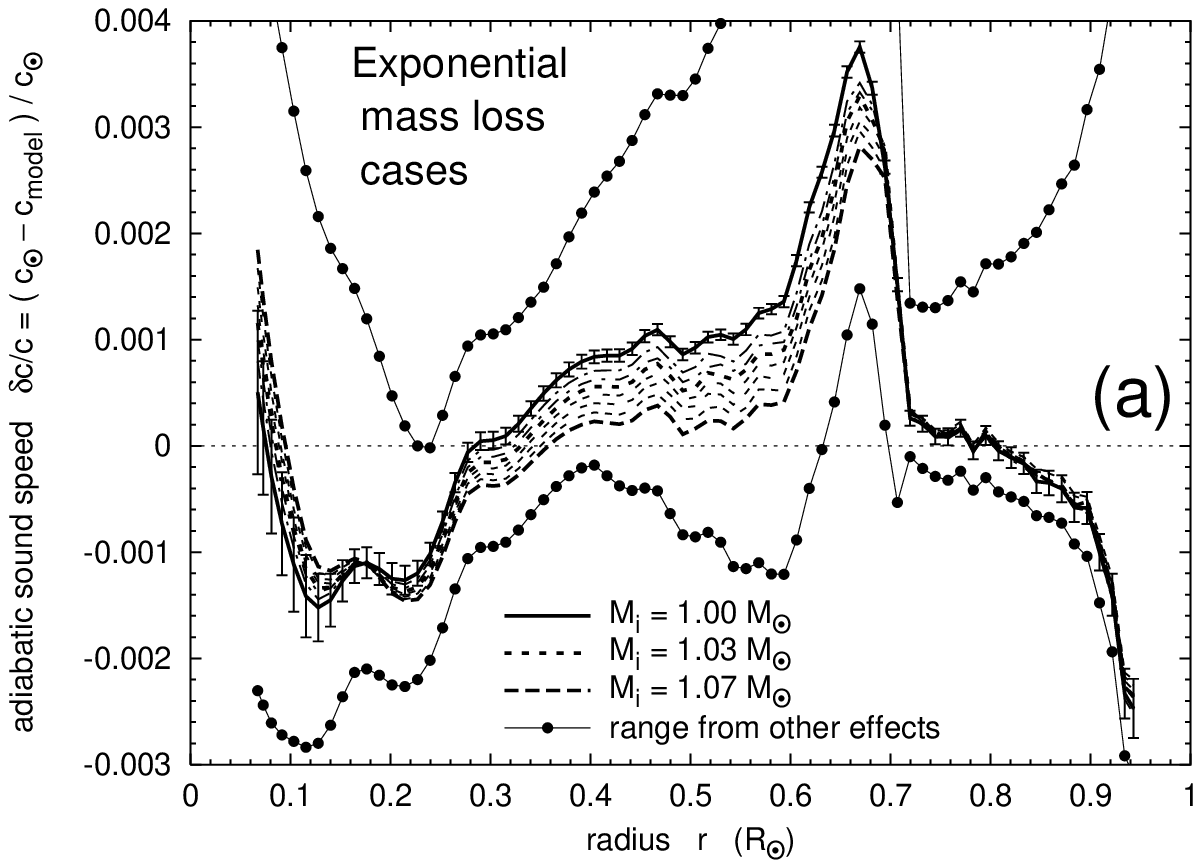}{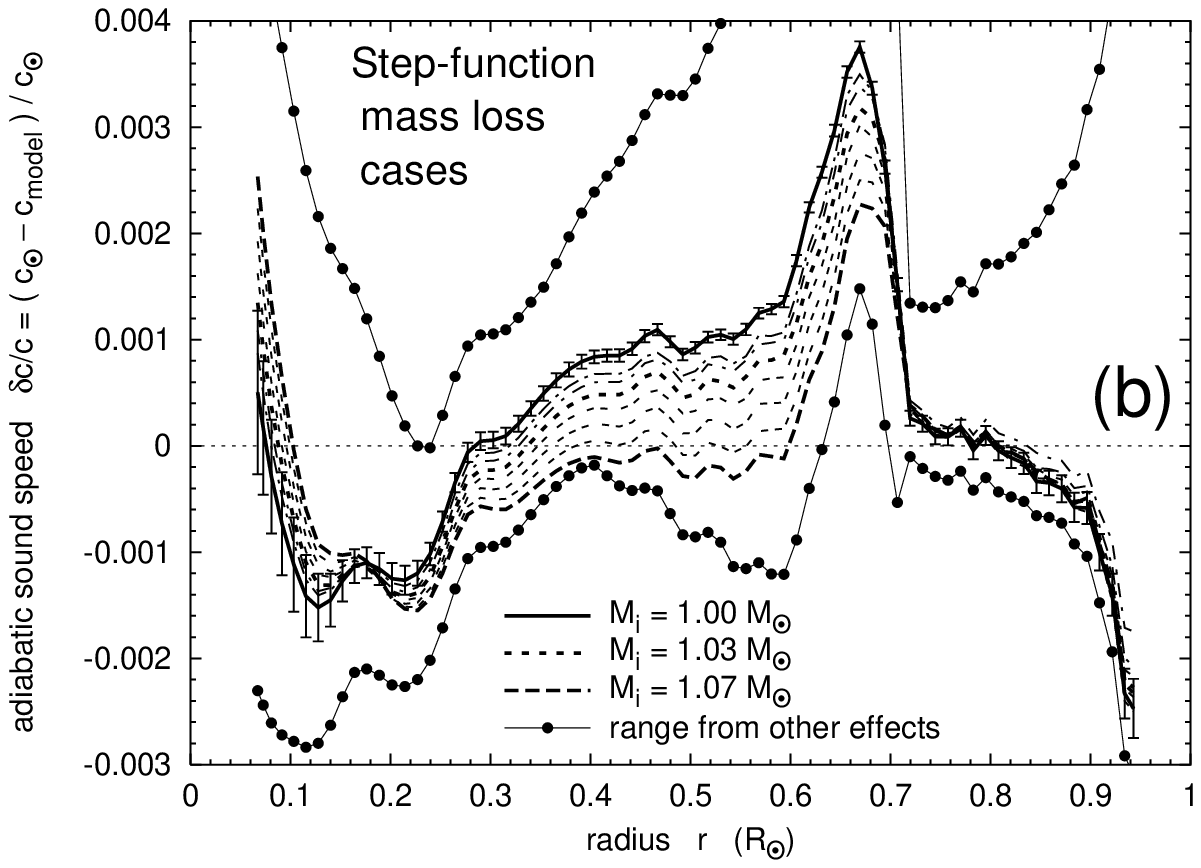}
  \epsscale{2.46}
 \plottwo{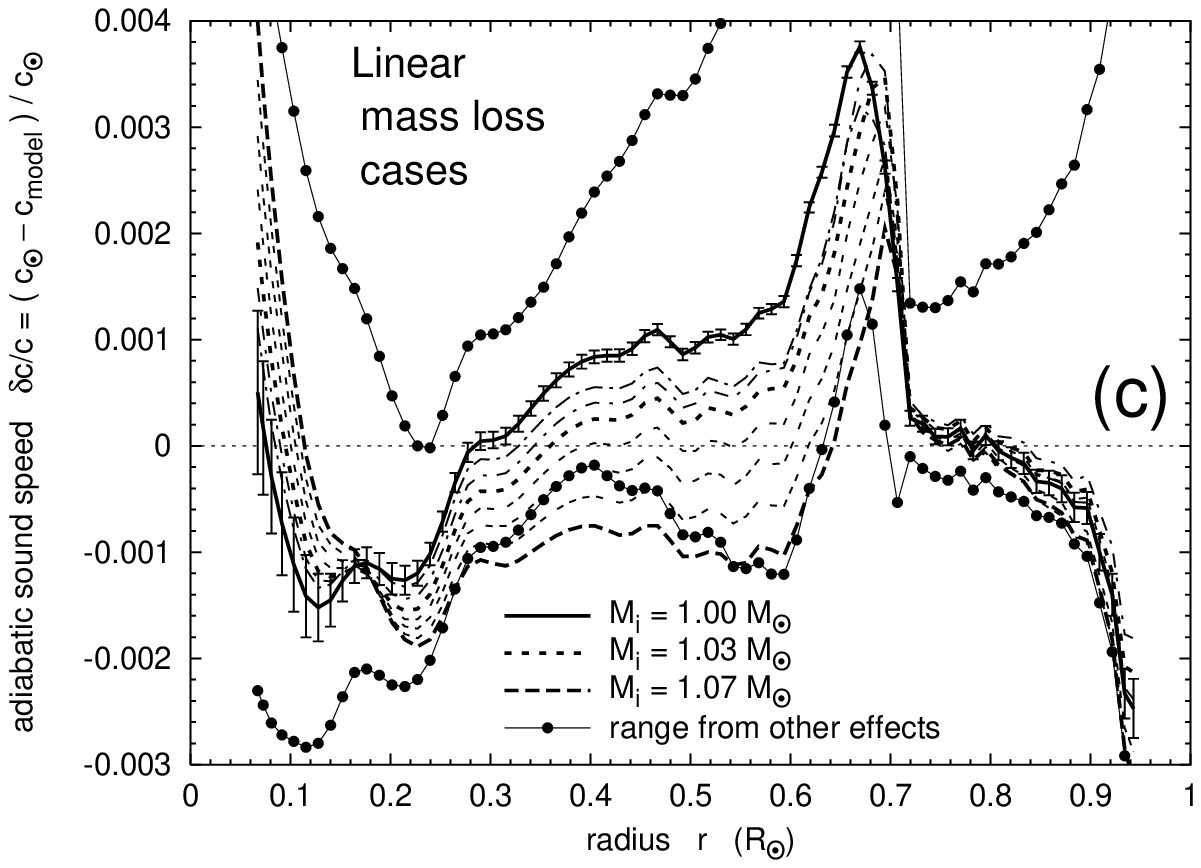}{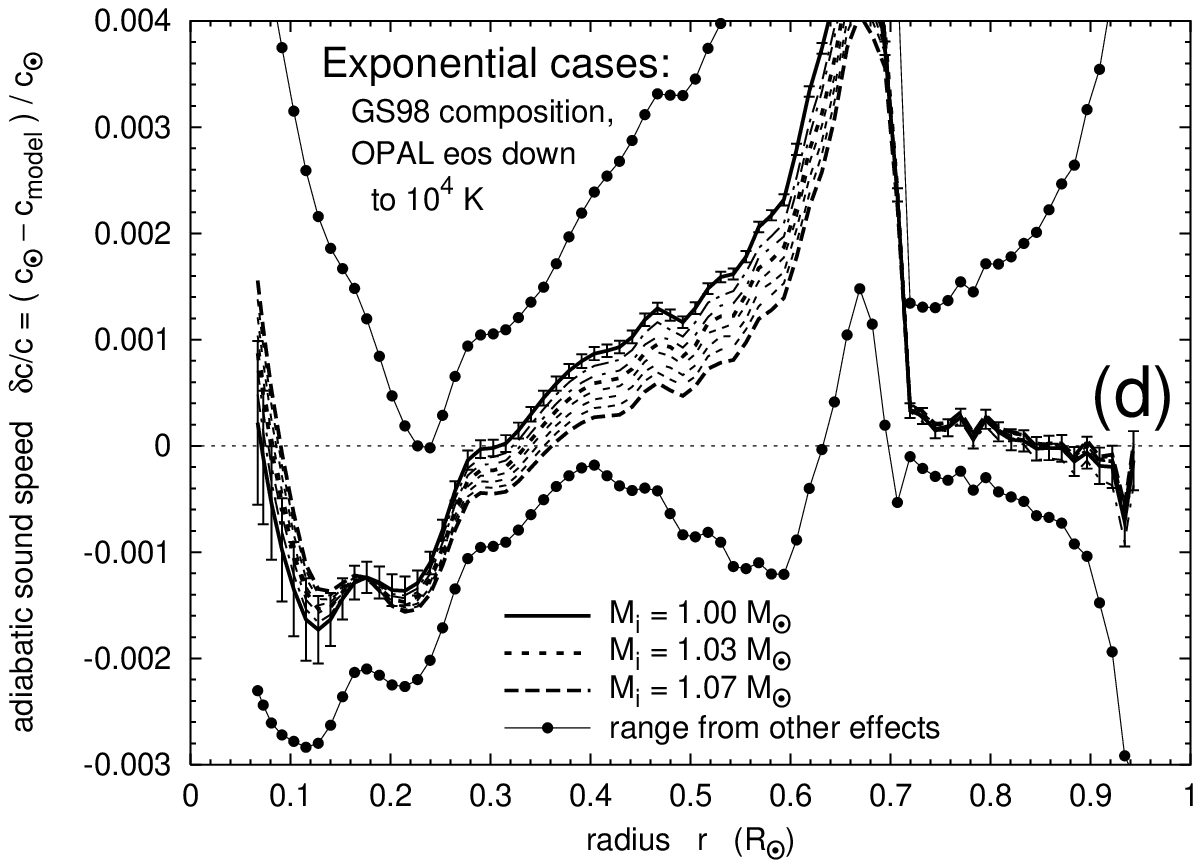}

\caption{The effects of mass loss on the adiabatic sound speed~$c$
for (a)~exponential, (b)~step-function, and (c)~linear mass loss cases;
(d)~shows exponential mass loss cases, as in~(a), except that it uses the
more recent ``GS98'' composition of
\citet{GreS98}
(with appropriate OPAL opacities) instead of the ``GN93'' composition of
\citet{GreN93},
and the OPAL equation of state is used at $\log\,T \gtrsim 4$ (rather than
at $\log\,\rho \gtrsim -2$, which corresponds to $\log\,T \gtrsim 5.5$).
The {\it heavy solid line\/} is the reference standard solar model
(no mass loss); the errorbars give the statistical error in the
inferred helioseismic profile.  Mass-losing cases are shown for initial
solar masses $M_i = 1.01 \; M_\odot$ ({\it close dot-dashed line\/}),
$M_i = 1.02 \; M_\odot$ ({\it wide dot-dashed line\/}),
$M_i = 1.03 \; M_\odot$ ({\it heavy dotted line\/}),
$M_i = 1.04 \; M_\odot$ ({\it double-dotted line\/}),
$M_i = 1.05 \; M_\odot$ ({\it triple-dotted line\/}),
$M_i = 1.06 \; M_\odot$ ({\it quadruple-dotted line\/}), and
$M_i = 1.07 \; M_\odot$ ({\it heavy dashed line\/}).
The {\it filled circles\/} (connected by {\it light solid line\/}) show
the range of variation in $\delta c/c$ found to arise from reasonable
variations in solar initial parameters, from our companion paper ``Our Sun~IV''
\citep{BS02}.}

 \label{fig:helio}

\end{figure}

\begin{figure}[t]
  \epsscale{1.11}
 \plottwo{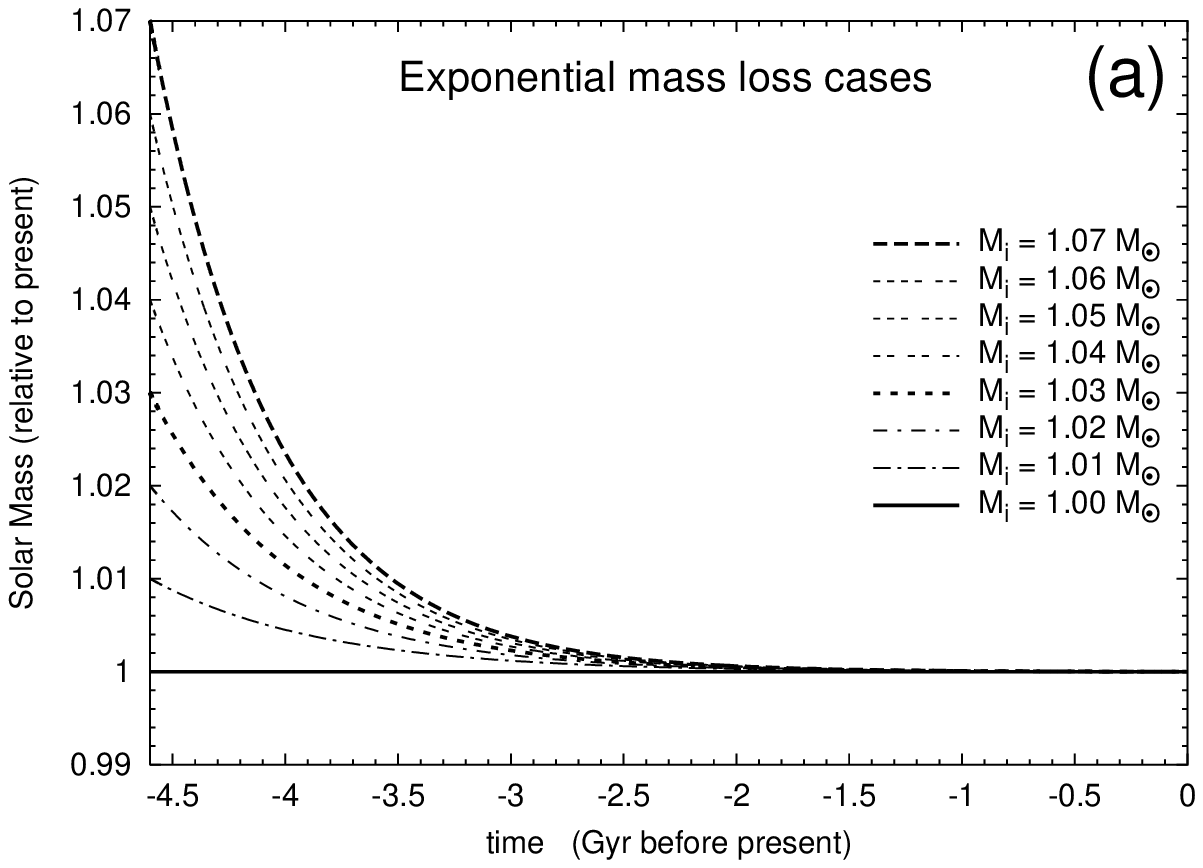}{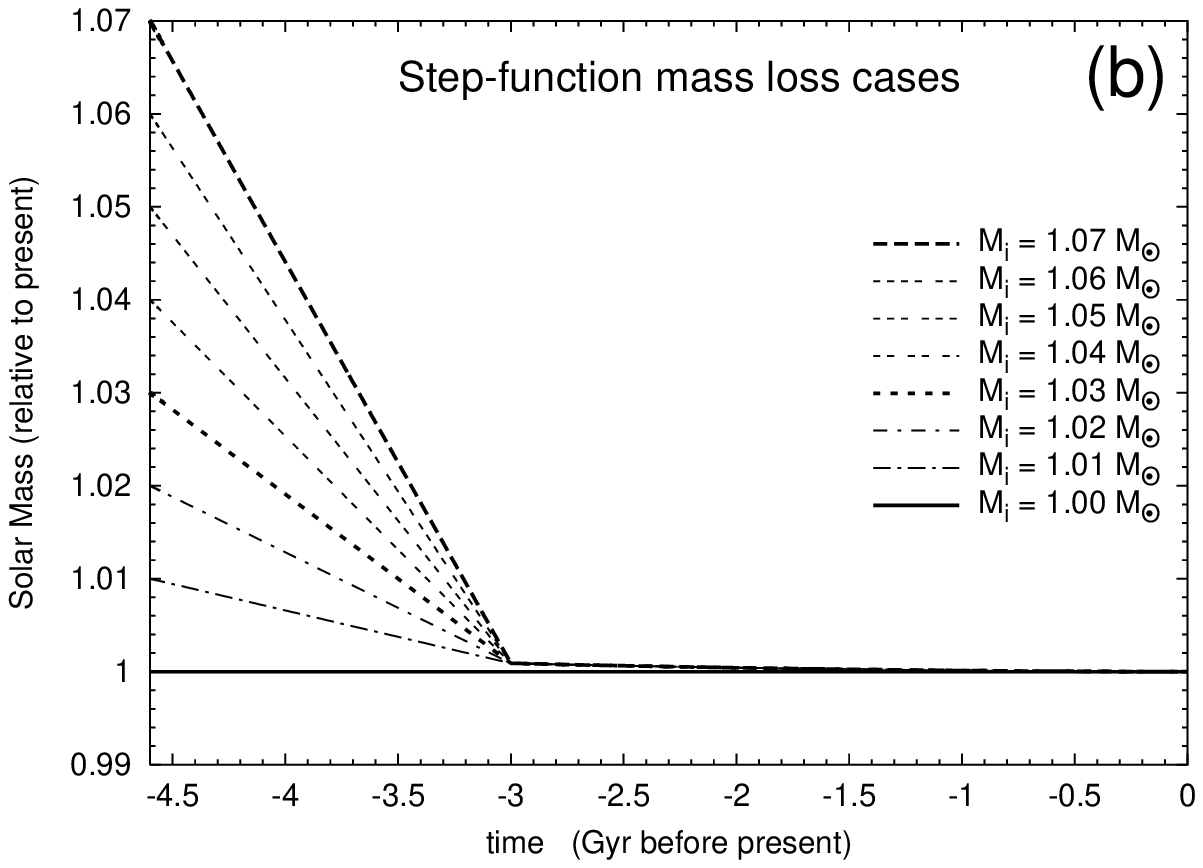}
 \plotone{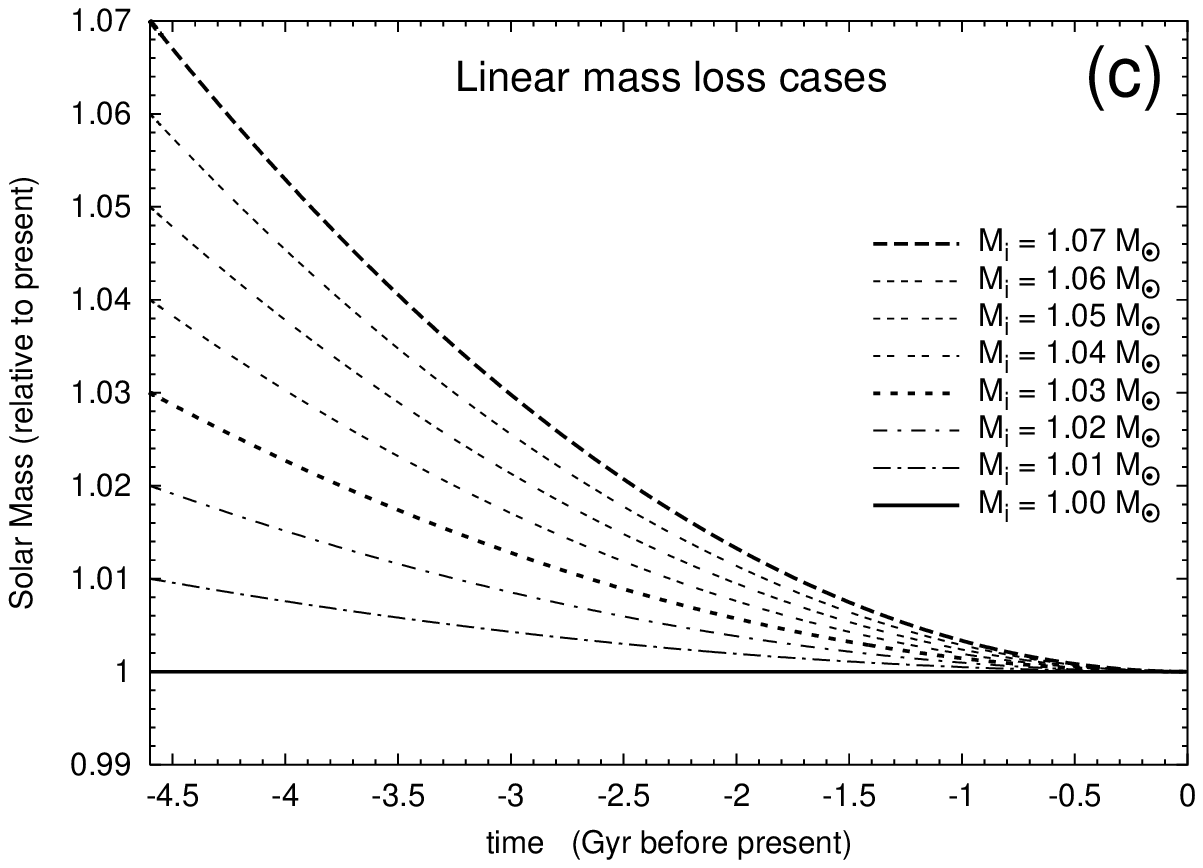}

\caption{Solar mass as a function of time for (a)~exponential,
(b)~step-function, and (c)~linear mass loss cases.}

 \label{fig:mass}

\end{figure}

We present in Figure~\ref{fig:helio}
profiles of the adiabatic sound speed differences
$\delta c / c \equiv ( c_\odot - c_{model} ) / c_\odot$;
profiles of the density differences
$\delta \rho / \rho \equiv ( \rho_\odot - \rho_{model} ) / \rho_\odot$
are available online\footnote{
\url[http://www.krl.caltech.edu/~aib/papdat.html]{http://www.krl.caltech.edu/${}^{\sim}$aib/papdat.html}
}.
Note that we use~``$\delta$'' to denote differences between the helioseismic
profile and one of our models, and ``$\Delta$'' to denote differences between
two of our models with different input parameters --- the ``$\delta$''~values
are the profiles plotted in Figure~\ref{fig:helio}, while the
``$\Delta$''~values
refer to the difference between one plotted curve and another.
Solar masses as a function of time for the corresponding cases are
presented in Figure~\ref{fig:mass}, and solar fluxes at the planets
(relative to their present values) are presented in Figure~\ref{fig:solarflux}.



Since the prominent peak at $r \sim 0.7 \; R_\odot$ results from
the neglect of rotational mixing, we did not require agreement
in this region between profiles from our theoretical models and
profiles inferred from the helioseismic observations.
Nor did we require agreement in core region, since the present helioseismic 
observations still result in large uncertainties in the inferred
profiles there.  On the other hand, we aimed for agreement in the regions
$0.1 \; R_\odot \lesssim r \lesssim 0.6 \; R_\odot$
and $0.72 \; R_\odot \lesssim r \lesssim 0.94 \; R_\odot$,
where disagreements
are due to imperfections in the input physics or uncertainties in
the observed solar parameters.
This is demonstrated by our variant models of
our companion paper ``Our Sun~IV''
\citep{BS02}
and of
\citet{MorPB97}
and
\citet{BasuPB00}.
Except around $r \sim 0.7 \; R_\odot$ and near the surface, the sound speed
profile in our standard solar models with the
\citet{GreN93}
composition ({\it heavy solid lines\/} in Fig.~\ref{fig:helio}a,~b,~c)
differ by $\delta c / c \lesssim 0.001$ from the helioseismic sound speed,
typical agreement for standard solar models with this composition
\citep[see, e.g.,][]{MorPB97,BasuPB00,BahPB01,TurckC+01b,NeuV+01a}.
However, if one updates to the more recent solar composition value of
\citet{GreS98},
there is a systematic shift in the sound speed profile in the interior
that worsens agreement with helioseismology by a factor of~$\sim 2$
({\it heavy solid line\/} in Fig.~\ref{fig:helio}d --- differences near
the solar surface are due to extending the use of the OPAL equation of
state to lower temperatures in this model).  This composition effect
was also demonstrated by solar models of
\citet{NeuV+01a}.
Even larger systematic effects are allowed by the quoted uncertainties in
the solar composition, and (to a lesser extent) the uncertainties in the
OPAL opacities, the $pp$~nuclear reaction rate, and the diffusion constants
for gravitational settling, as shown in detail in
our companion paper ``Our Sun~IV''
\citep{BS02}.
In Figure~\ref{fig:helio}, the {\it filled circles\/} (connected by
{\it light solid lines\/}) indicate the extremes of the range of
$\delta c / c$ profiles found in
\citet{BS02}
for solar models with ``reasonable'' variations in these input parameters
(i.e., variations allowed by the quoted uncertainties).  In the comvective
envelope, the largest effects on the sound speed profile arises from the
uncertainties in the observed solar radius and in the equation
of state; in the bulk of the solar interior, the largest effects
arise from uncertainties in the observed solar surface composition;
in the core, the largest effect arises from the uncertainty in the
$pp$~nuclear reaction rate
\citep{BS02}.

Figure~\ref{fig:helio}a,~d demonstrates that all of our ``exponential''
mass loss models agree better with the helioseismic observations
than the standard solar model, which has no mass loss
(i.e., they lie closer to the zero line of perfect agreement).
However, this improvement is clearly not significant: even for our most
extreme ``exponential'' mass loss case (with $M_i = 1.07 \; M_\odot$),
the effect is only about one third as large as the maximum effect allowed by
variations in other input solar parameters (also shown in
Fig.~\ref{fig:helio}).

The rms sound speed differences
provide a numerical measure of the extent of the above
agreement between a given theoretical model and the profile inferred from
helioseismic observations; these rms values for each of our mass-losing
cases are given in Table~\ref{tab:results}
(for completeness, the fine-zoned cases are presented as well as the
coarse-zoned ones, although the latter proved to be quite accurate
enough, as discussed in \S~\ref{ssec:inputs}).  The differences
rms$\{\delta c / c\}$ relative to the helioseismic sound speed profile
are shown both for the entire Sun and for the region $r < 0.6 \; R_\odot$
(where the latter region excludes the peak near $r \sim 0.7 \; R_\odot$
that results from neglect of rotation-induced mixing, and the envelope
which is sensitive mostly to the solar radius and the equation of state);
unlike the sound speed, the density profile does not respond in a localized
way to local variations in the solar models, so only the
rms$\{\delta \rho / \rho\}$ over the entire Sun is shown.
The differences rms$\{\Delta c / c\}$ (and rms$\{\Delta \rho / \rho\}$)
relative to the corresponding theoretical standard (non-mass-losing)
solar model are shown for the entire Sun, since they measure the
differential effects of mass loss.

Table~\ref{tab:results} shows that the ``exponential''
$M_i = 1.07 \; M_\odot$ cases have rms$\{\delta c / c\}$ 
errors about 30\% smaller than the corresponding standard solar
models, but again this difference is not significant compared to the
large variations that can arise from uncertainties in the other input
solar parameters (shown in the last line of Table~\ref{tab:results}).
This is perhaps most easily seen by comparing the
relative rms measures: the ``exponential'' cases differ by
rms$\{\Delta c / c\} \lesssim 0.0007$ from their corresponding standard
solar model, while the shifts in the sound speed profile from
uncertainties in other input parameters can be as large as
rms$\{\Delta c / c\} \sim 0.0018$, nearly three times as large.

In summary, none of our ``exponential'' mass loss cases with initial
solar masses $M_i \le 1.07 \; M_\odot$ can be ruled out by
helioseismological observations; in fact, all other things being equal,
these helioseismological observations slightly favor the ``exponential''
mass loss case with highest of our initial solar
masses ($M_i = 1.07 \; M_\odot$), although this is not statistically
significant, as shown by Figure~\ref{fig:helio}a,~d
and the rms values in Table~\ref{tab:results}.

Detailed investigation of the effects of the semi-empirical mass loss law
$\dot M \propto t^{-2.00\pm0.52}$ recently presented by
\citet{Wood+02}
and discussed in \S~\ref{sssec:mdotobs} above
must wait for a future paper
\citetext{A.~I.~Boothroyd \& \hbox{I.-J.}~Sackmann, in preparation}.
However, the total amount of mass lost by the Sun would be comparable to
or smaller than that in our ``exponential'' mass loss case, and the
evolution of the mass loss rate as a function of time would be not too
dissimilar (in both cases, most of the mass loss occurs at quite early
times).  Thus a $t^{-2}$ mass loss case should yield an effect on the
solar sound speed profile similar to (or smaller than) that of
our ``exponential'' mass loss cases of Figure~\ref{fig:helio}a,~d.



Figure~\ref{fig:helio}b presents our results for our ``step-function''
mass loss models; recall that this type of mass loss was chosen as the
most extreme possibility that remains consistent with all the observational
mass loss constraints (see \S~\ref{ssec:mdotused}).
Since the Sun remains more massive for a longer period in this
case than for
the ``exponential'' mass loss cases (compare Fig.~\ref{fig:mass}b with
Fig.~\ref{fig:mass}a), there is more of an impact on the Sun's
internal structure (compare Fig.~\ref{fig:helio}b with
Fig.~\ref{fig:helio}a).  Again, the mass-losing models with higher
initial masses fit slightly (but not significantly) better than the
standard non-mass-losing model, as shown by Figure~\ref{fig:helio}b
and the rms values in Table~\ref{tab:results}.
Even our most extreme ``step-function'' case ($M_i = 1.07 \; M_\odot$)
differs from the standard solar model by an
rms$\{\Delta c / c\} \lesssim 0.0010$, as compared to 0.0018 from
other uncertainties.
As in the ``exponential'' mass loss
case, none of the ``step-function'' mass loss cases can be ruled out by
helioseismic observations.

Figure~\ref{fig:helio}c presents our results for the radical ``linear''
mass loss models; recall that these models violate the solar wind
constraints from the lunar rock observations by about an order of magnitude
(see \S~\ref{ssec:mdotused}).
Since the Sun remains remains more massive for a very long time
compared to the other mass loss cases (compare Fig.~\ref{fig:mass}c with
Fig.~\ref{fig:mass}a,~b), there is even more of an impact on the Sun's
internal structure (compare Fig.~\ref{fig:helio}c with
Fig.~\ref{fig:helio}a,~b).
Intermediate $M_i$ values have the lowest rms$\{\Delta c / c\}$ values,
though again none of the differences are significant; but the most
extreme ``linear'' $M_i = 1.07 \; M_\odot$ differs from the standard
solar model by rms$\{\Delta c / c\} \sim 0.0016$, nearly as large as
the largest effect of 0.0018 from other uncertainties.
Even these radical ``linear'' mass loss cases cannot yet be ruled out
by comparisons with the helioseismic observations.

\subsubsection{Position of Convection and Surface Helium Abundance}
\label{sssec:RceYe}

Helioseismic observations measure the position of the base of the
Sun's convective envelope, namely $R_{ce} = 0.713 \pm 0.001 \; R_\odot$
\citep{BasuA97},
and the surface helium abundance, namely, a mass
fraction~$Y_e \approx 0.245 \pm 0.005$
\citep[see discussion in our companion paper ``Our Sun~IV'':][]{BS02}.
The values of $R_{ce}$ and~$Y_e$ for
both our reference standard solar model and for all of our mass-losing
models are shown in Table~\ref{tab:results}.  The mass-losing cases
all have values of $R_{ce}$ and~$Y_e$ very close to those of the
standard solar model, all of them being consistent with the
helioseismic observations.

\subsection{Other Effects of Modest Mass Loss}  \label{ssec:mdotother}

\subsubsection{Solar Lithium Depletion}  \label{sssec:lithium}

The lithium depletion in a main sequence star, relative to its initial
lithium abundance, can result from three different causes.  (1)~There can
be significant lithium depletion from pre-main-sequence lithium
burning at early times, when the convective envelope reaches deep into
the star.  (2)~Rotationally induced mixing on the main sequence can
transport lithium down from the convective envelope to regions hot
enough for lithium burning.  (3)~Mass loss on the main sequence
can cause the convective envelope to move inwards and
engulf lithium-depleted regions.

The Sun's initial lithium abundance
is assumed to be equal to the meteoritic abundance, and the depletion
factor~$f_{\rm Li}$ is obtained by comparing this initial abundance
with the present observed solar photospheric lithium abundance.
This observed solar lithium depletion factor is $f_{\rm Li} = 160 \pm 40$
\citep{GreS98}.
For pre-main-sequence lithium depletion, our reference standard solar
model predicts a lithium depletion factor of $f_{\rm Li} \sim 24$,
although this is very sensitive to the solar metallicity (i.e.,
$Z/X$ value, as well as uncertainties in diffusion) and to the choice of
low-temperature molecular opacities;
values of $f_{\rm Li}$ from~11 to~70 can be obtained
\cite[see our companion paper ``Our Sun~IV'':][]{BS02}.
For rotation-induced main sequence lithium depletion, there is no
theoretical prediction; instead, the observed main sequence lithium
depletion is used to constrain the free parameters in the theoretical
treatment
\citep[see, e.g.,][]{Schat77,LebM87,Pins+89,CharVZ92,Rich+96}.
For main sequence mass loss, the extent of main sequence lithium
depletion depends primarily on the initial solar mass, and only
weakly on the timescale of mass loss.
\citet{BSF91}
used the observed solar lithium depletion to obtain a limit on
solar main sequence mass loss, finding that the maximum mass loss
allowed was~$0.1 \; M_\odot$ (i.e., a maximum initial solar mass of
$M_i \approx 1.1 \; M_\odot$).  However, as discussed in \S~\ref{sec:intro},
such an extreme mass loss case violates the constraint from the
requirement that the early Earth not lose its water via a moist
greenhouse effect, which would occur for $M_i > 1.07 \; M_\odot$.
This constraint is based on a cloud-free climate model; a very slight
increase in~$M_i$ might be allowed if clouds were taken into account.

As the Sun's initial mass is increased above~$1 \; M_\odot$, there
are two competing effects.  Higher initial masses have less
pre-main-sequence lithium depletion; on the other hand, the higher
the initial mass, the more mass loss has to take place, and thus
the more lithium depletion takes place on the main sequence (as the
convective envelope sheds lithium-rich material from the surface and
engulfs lithium-depleted material from below).  As may be seen from
Table~\ref{tab:results},
for initial solar masses in the range
$1.01 \; M_\odot \le M_i \lesssim 1.04 \; M_\odot$, the
first of these effects dominates, and the total lithium depletion
is slightly less than in the standard (non-mass-losing) model;
slightly stronger rotation-induced mixing would be required in order
to account for the observed lithium depletion.  For masses
$M_i \gtrsim 1.05 \; M_\odot$, the mass loss dominates; however, even for
our most extreme mass loss cases ($M_i = 1.07 \; M_\odot$), the total
lithium depletion (from pre-main-sequence burning plus mass loss effects)
is only $f_{\rm Li} \sim 30 - 50$.
This is at most factor of~2 more lithium depletion than
in the standard (non-mass-losing) model with $f_{\rm Li} \sim 24$.

The modest amount of mass loss considered here
($\Delta M \le 0.07\;M_\odot$) has only a minor effect on the extent
of solar lithium depletion --- adding even the maximum possible mass loss
only increases the lithium depletion by a factor of~2, relative to the
depletion on the pre-main-sequence.
This effect is smaller than the factor-of-2.5 effects on pre-main-sequence
lithium depletion caused by uncertainties in other physical parameters, as
discussed above and in our companion paper ``Our Sun~IV''
\citep{BS02}.
In these mass-losing models, rotational mixing would still be required, to
account for the majority of the Sun's main-sequence lithium depletion;
therefore the arguments of
\citet{Swen+94}
(who showed that mass loss could not be responsible for the majority of the
observed Hyades lithium depletion) are not applicable,
as discussed in \S~\ref{sssec:hyrest}.
The observed solar lithium depletion thus cannot be used to constrain
these mass-losing solar models.

\subsubsection{Solar Beryllium Depletion} \label{sssec:beryllium}

The observed solar beryllium abundance is
$\log\,\varepsilon(\iso9{Be}) = 1.40 \pm 0.09$,
consistent with no depletion relative to the meteoritic value of
$\log\,\varepsilon(\iso9{Be}) = 1.42 \pm 0.04$.
The uncertainties given for these values imply that
solar beryllium cannot have been depleted by more than a
factor of~2 (3-$\sigma$ upper limit).
A standard solar model has negligible beryllium depletion ($\sim 1$\%);
our mass-losing solar models predict slightly larger depletions, but
are still all consistent with the observational limit.
The most extreme of the ``exponential'' mass loss cases
($M_i = 1.07\;M_\odot$) depleted beryllium by a relatively small
amount (a factor of~1.17).  Even the most extreme of the
``step-function'' and ``linear'' mass loss cases yielded only
$\sim 2$-$\sigma$ beryllium depletion factors (of 1.53 and~1.63,
respectively); the $M_i = 1.04\;M_\odot$ cases depleted beryllium
by negligible amounts, less than~3\%.  In other words, all the mass losing
cases deplete beryllium by significantly less than the observational
upper limit of a factor of~2.

\subsubsection{Neutrino Fluxes} \label{sssec:neutrinos}

As may be seen from Table~\ref{tab:results}, the modest mass loss
considered here has almost no effect on the predicted solar neutrino
fluxes.  Variations are at most a few percent in the predicted \iso8B flux
and in the predicted capture rate for the chlorine experiment (as compared
to uncertainties of $\sim 30$\% from other causes), and less
than a percent in the predicted capture rate for the gallium experiment
(as compared to uncertainties of at least several percent from other
causes) --- the other sources of uncertainty in neutrino fluxes are
discussed elsewhere
\citep[see, e.g.,][]{BahPW95,BahPB01,BS02}.

\subsection{The Young Earth and the Solar Flux} \label{ssec:earth}

At present, an airless, rapidly-rotating body at Earth's orbit would
have a temperature of~$-18^\circ\,$C (255~K),
if it had Earth's present albedo and emissivity
\citep{SagC97},
but the Earth's present
mean surface temperature is observed to be~$+15^\circ\,$C (288~K)\hbox{}.
In other words, at present the greenhouse effect raises Earth's surface
temperature by~$33^\circ\,$C\hbox{}.
The main greenhouse gases in the Earth's atmosphere are
CO$_2$ and H$_2$O.  If the atmospheric CO$_2$ abundance were
{\it constant\/} (at its present value), and the H$_2$O abundance were
determined by its equilibrium vapor pressure, then 2~Gyr ago the
Earth's surface temperature would have been below $0^\circ\,$C
\citep{SagM72,Sagan77,Poll79}.
If the early Earth's surface temperature were below the freezing point of
water, extensive glaciation would be expected; such glaciation would raise
the Earth's albedo, delaying the time when the surface temperature reached
$0^\circ\,$C\hbox{}.  In other words, one would expect Earth to be fully
glaciated as recently as 1~Gyr in the past
\citep{North75,WangS80}.

On the other hand, a number of independent observations
indicate that the Earth was at least
warm enough for liquid water to exist as far back as 4~Gyr ago.
Sedimentary rocks, which are laid down under water, have been dated to at
least 4~Gyr ago
\citep{Bow89,Nut+84}.
Liquid water is necessary to explain the existence of the widespread
microorganisms whose fossils are found
in rocks dated as far back as 3.8~Gyr ago
\citep{CogH84,Moj+96,EilMA97}.
Tidal or intertidal stromatolite fossils have been dated to $\sim 3.5$~Gyr
ago, alluvial detrital uraninite grains as far back as 3~Gyr, and
turbidites and ripple marks have been dated as far back as 3.5~Gyr
\citep{Erik82}.

In fact, there is evidence not only that liquid water existed on the early
Earth, but also that Earth was considerably warmer in the past than it
is today.
To start with, there is no evidence of glaciation before 2.7~Gyr ago
\citep{Kast89},
and it has been suggested that tillites prior to 2~Gyr ago are actually
due to impacts rather than glaciers
\citep{OberMA93}.
Archaeobacteria exhibit extreme thermophilic trends
\citep{Woese87}.
High ocean temperatures of $\sim 40^\circ\,$C in the period 2.6 to 3.5~Gyr
ago are suggested by sulphur isotope measurements
\citep{OhmF87}.
Average surface temperatures of tens of degrees Celsius in the period
2.5 to 3.5~Gyr ago are indicated by deuterium to \iso{18}O ratios
\citep{KnauE76}.
Temperatures as high as $80^\circ\,$C in the period $\sim 3.8$~Gyr ago
are suggested by differences in \iso{18}O isotopic data between
coexisting cherts and phosphates
\citep{KarE86},
although the results are subject to interpretation.

The above ``weak Sun paradox'', of a faint young Sun and a young Earth warm
enough for liquid water,
has traditionally been explained by invoking a much stronger greenhouse
effect for ancient Earth.  This would be
driven by extremely high (i.e., {\it non-constant\/}) CO$_2$
concentrations in Earth's early atmosphere, with partial CO$_2$ pressures
between 0.2 and 10~bars 4.5~Gyr ago
\citep{Poll79,KuhnK83,KastA86,Kast87}.
Qualitatively, high CO$_2$
concentrations can be justified on the basis of theoretical feedback
mechanisms linking mineral dissolution to liquid water and thus to
atmospheric CO$_2$
\citep{WalHK81}.
Although such massive amounts of CO$_2$ in the Earth's early atmosphere
are a possible solution to the ``weak Sun paradox'', they
are not mandated;
there is little experimental evidence available on which to base a choice
of CO$_2$ concentration
\citep{Can+83,KuhnWM89}.
Indeed, very high concentrations may prove to be inconsistent with derived
weathering rates
\citep{HolLM86}.
The recent work of
\citet{RyeKH95}
places an upper limit of 0.04~bar on the partial pressure of CO$_2$ in the
period from 2.75 to 2.2~Gyr ago, based on the absence of siderite in
paleosols; earlier work by
\citet{HollZ86}
estimated a partial pressure of CO$_2$ of 0.004~bar, but with the same
upper limit of 0.04~bar.
For Earth surface temperatures between 5 and~20$^\circ$C during that period,
climate models predict a partial pressure of CO$_2$ between 0.03 and 0.3~bar
\citep{Kast87},
barely consistent with the upper limit of 0.04~bar.
In other words, there is little evidence for a strong CO$_2$ greenhouse
effect on ancient Earth.

Actual measurements of CO$_2$ abundances are available
only for relatively recent times, i.e., only for the last $\sim 0.45$~Gyr
\citep[see, e.g.,][]{CroB01,Ret01}.
These latter measurements show major variations in the CO$_2$
abundance over the past 0.45~Gyr.  The lowest values are comparable to
the present-day CO$_2$ abundance of about 350~ppmV, namely, 0.00035~bar
(or the pre-industrial-age value of $\sim 300$~ppmV --- note that ``ppmV''
refers to parts per million by volume);
the highest values measured over the last 0.45~Gyr are
$\sim 5000$~ppmV\hbox{}.  However, from these measurements, there is no
clear evidence of a long-term trend of higher CO$_2$ abundances in the
relatively recent past (i.e., the last 0.4~Gyr).

A non-CO$_2$ greenhouse has been suggested for the early Earth
\citep{Kast82,Love88}.
Recently,
\citet{SagC97}
calculated that a strong greenhouse contribution from ammonia was possible,
if a concentration of $\rm [NH_3] \sim 10^{-5}$ could be maintained.
Normally, the ammonia would be photodissociated by solar UV flux on a
timescale of 10 years.  They
pointed out that ammonia could be shielded from the UV radiation by
high-altitude organic solids produced from photolysis of methane
--- a photochemical smog, similar to that observed
in the upper atmosphere of Titan
\citep{RagP80}.
However, the ammonia lifetime depends sensitively
on two parameters, the fraction~$f$ of the methane irradiation
products that are organic solids, and the sedimentation
timescale~$t$ of the smog;
\citet{SagC97}
take as reasonable values $f \gtrsim 0.1$ and
0.5~yr${} \lesssim t \lesssim 3$~yr.  For $f \gtrsim 0.5$,
or $t \sim 3$~yr, the ammonia lifetime is long enough for the required
ammonia concentration to be maintained, given a reasonable amount of
resupply.  However, for $f \sim 1$ and $t \lesssim 1$~yr, 
the ammonia lifetime is less than 200~yr, which would require excessively
large amounts of ammonia production to maintain the required ammonia
greenhouse effect.  
\citet{SagC97}
also note that an atmosphere rich in N$_2$, with minor CO$_2$ and CH$_4$
components, could have adequate self-shielding of NH$_3$ from
photodissociation (allowing an ammonia greenhouse),
only as long as the ratio CH$_4$/CO$_2 \gtrsim 1$ was maintained.
In other words, for the early Earth, a smog-shielded ammonia greenhouse
is a viable solution to the ``weak Sun paradox'' under certain conditions,
but fails under others.

As discussed above, it is not clear whether the greenhouse effect could
suffice to warm the early Earth.  A bright young Sun, with stronger
illumination of the young Earth than from the standard solar model,
would require a less extreme early greenhouse effect to prevent the
early Earth from freezing over.

\subsection{The Young Mars and the Solar Flux} \label{ssec:mars}

For Mars, there are also indications of higher surface temperatures in
the past, that are even harder to explain by a greenhouse effect.
There is evidence of large scale flow of liquid water $\sim 3.8$~Gyr
ago, from the drainage channels and valley networks visible on the
heavily cratered ancient surface of Mars
\citep{Poll+87,Carr96};
there is some evidence for lakes
\citep{GoldS91,Park+93},
and possibly even oceans 3 to 4~Gyr ago
\citep{Schaef90,Baker+91}.
Even if the channels were formed by subsurface
sapping of groundwater, Martian surface temperatures significantly
higher than today would have been required for liquid water to be
present near the surface; if the suggested evidence of glacial markings
were confirmed, this would require temperatures high enough for
precipitation to occur
\citep{Whit+95}.

\begin{figure}[t]
  \epsscale{1.11}
 \plottwo{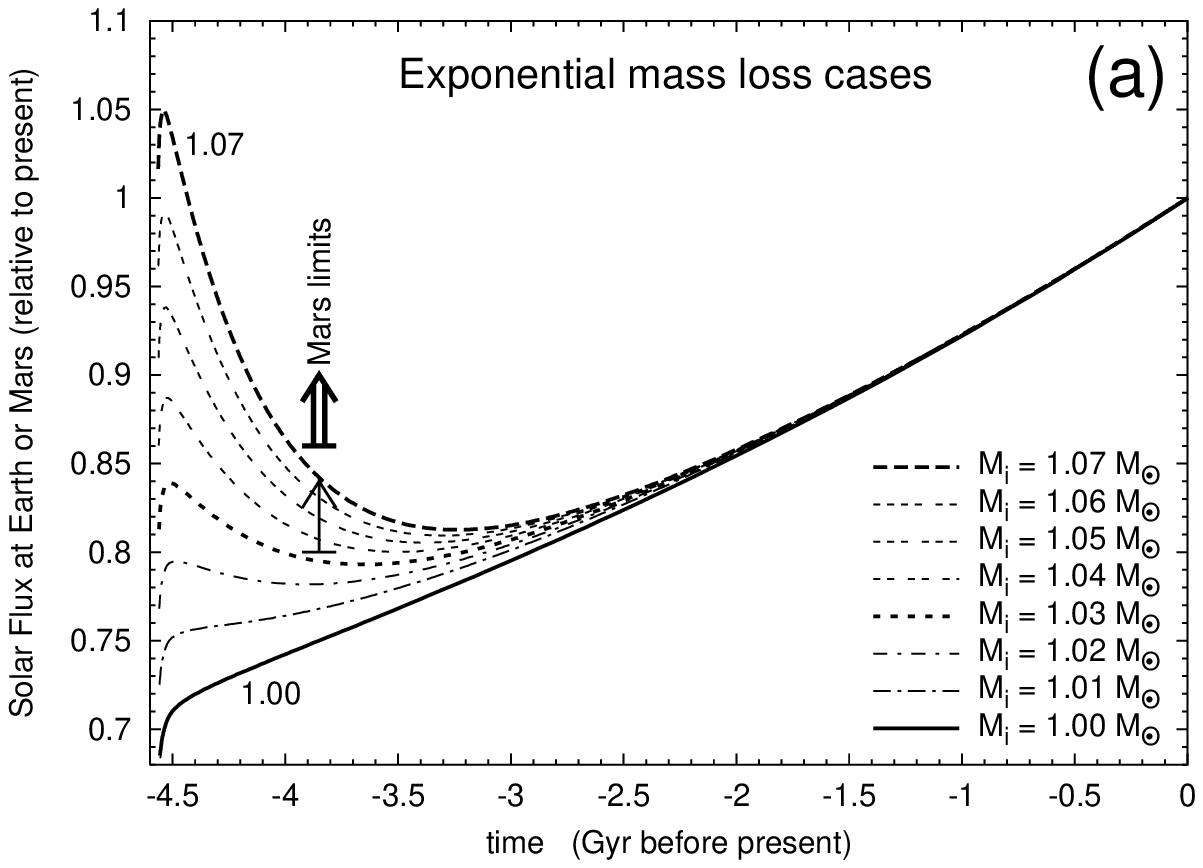}{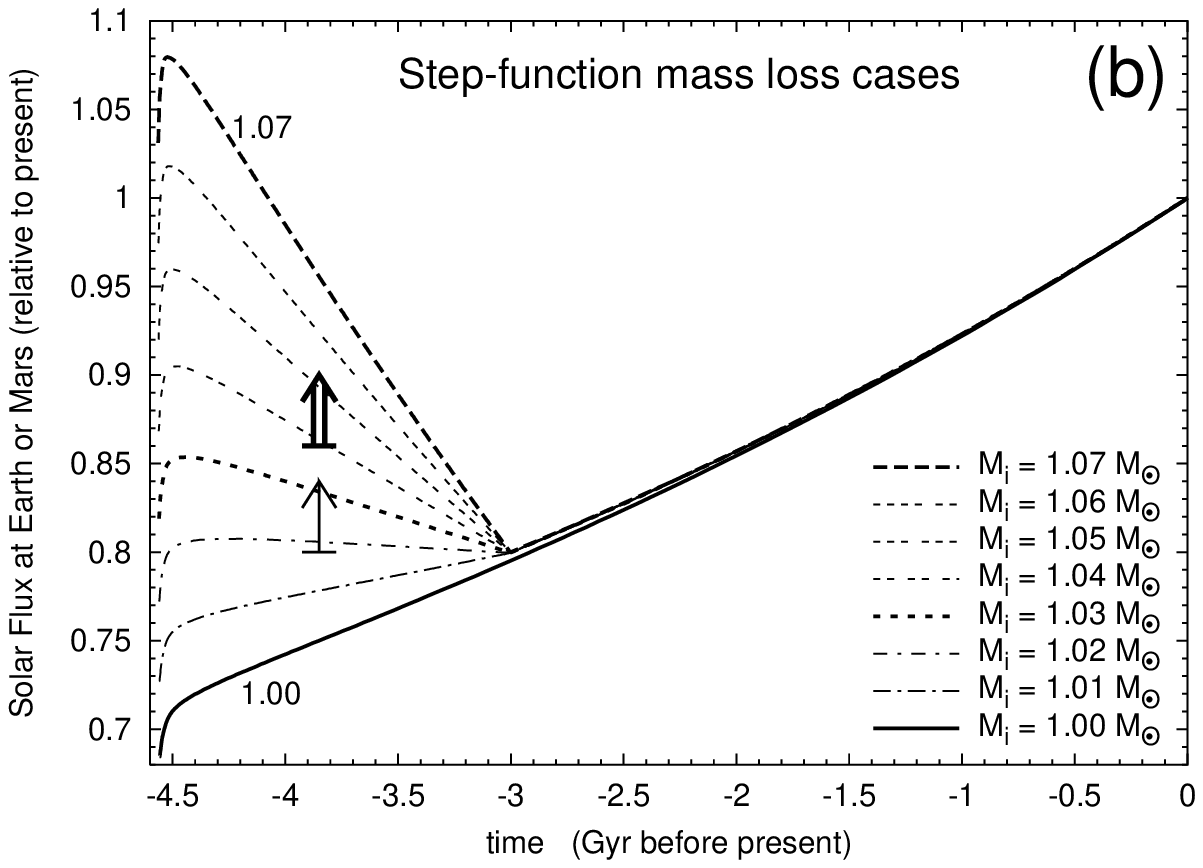}
 \plotone{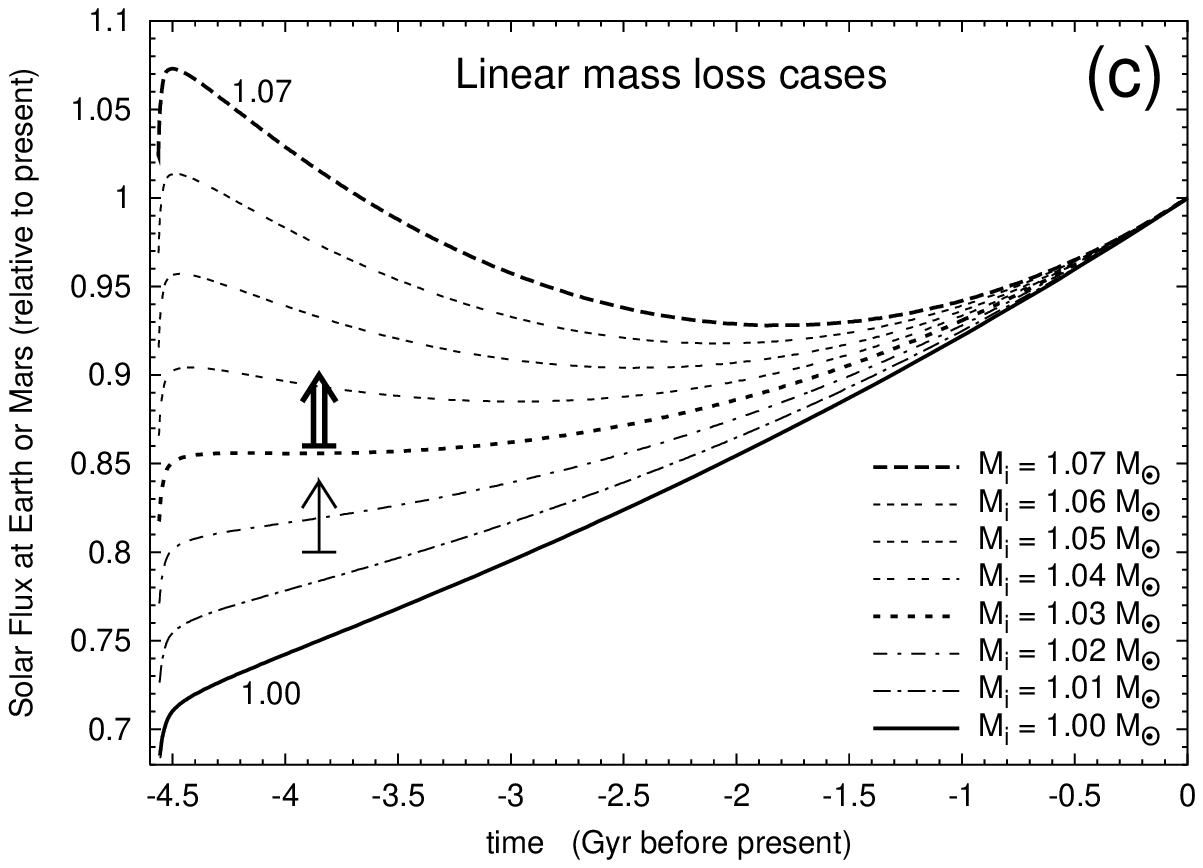}

\caption{Solar flux at the planets as a function of time (relative to
the present flux), for (a)~exponential, (b)~step-function, and (c)~linear
mass loss cases.  Heavy double arrows give the lower flux limit of
\citet{Kast91}
for the presence of water on early Mars; light single arrows give his
extreme lower flux limit (for a model with an unrealistically low
Martian surface albedo).}

 \label{fig:solarflux}

\end{figure}

\citet{Kast91}
demonstrated that there is an upper limit to the greenhouse
warming of Mars that is possible from CO$_2$.  He showed that the
maximum possible greenhouse warming occurs at a Martian surface
CO$_2$ pressure of 5~bars --- with more CO$_2$, the added greenhouse
warming is outweighed by the cooling effects of
increased CO$_2$ condensation in the upper Martian atmosphere.
He demonstrated that the
requirement of liquid water on Mars, i.e., a surface temperature of at
least~$273^\circ\;$K, demands a solar flux value $S \ge 0.86$ (where $S$ is the
solar flux relative to its present value), even with the most favorable
greenhouse warming case of a CO$_2$ pressure of 5~bars.  At 3.8~Gyr
ago, when liquid water is thought to have existed on Mars, the standard
solar model yields a value of $S = 0.75$ (see Fig.~\ref{fig:solarflux}),
totally insufficient relative to Kasting's minimum value of~0.86.
\citet{Kast91}
states that the uncertainties in his Martian climate model might
push the limiting value of~$S$ from 0.86 down to~0.80 (albeit for
an unreasonably low Martian albedo), but even this lower $S$ requirement
is incompatible with the standard solar model.  The standard solar model
does not reach this extreme limit of $S = 0.80$ until 2.9~Gyr ago,
and reaches $S = 0.86$ later still, at 1.9~Gyr ago
(see Fig.~\ref{fig:solarflux}) --- in either case,
far too late to account for liquid water on Mars 3.8~Gyr ago.
With CO$_2$ pressures either lower or higher than 5~bar,
\citet{Kast91}
shows that even higher solar flux values would be required to yield
liquid water.  He
presents the solar flux required to obtain liquid water as a function
of the CO$_2$ pressure: e.g., a pressure of 3~bar would require
$S \gtrsim 0.92$, and a pressure of 10~bar would require
$S \gtrsim 0.90$.  The standard solar model is totally incapable of
yielding such high fluxes 3.8~Gyr ago (see Fig.~\ref{fig:solarflux}).
In other words, for a standard solar model, CO$_2$ greenhouse warming
cannot under any circumstances yield liquid water on early Mars.

A preliminary study by
\citet{YungNG97}
suggested the possibility that relatively small amounts of atmospheric SO$_2$
($\sim 10^{-7}\;$bar) might have served as a powerful source of heating in
the upper atmosphere of early Mars (due to its strong absorption in the
near~UV), which might have been sufficient to prevent the condensation
of CO$_2$ (though they noted that SO$_2$ in the presence of liquid water
would produce H$_2$SO$_4$, which would lead to some countervailing cooling).
\citet{MurB98}
pointed out a more important effect of the H$_2$SO$_4$:
even $\sim 10^{-11}\;$bar of SO$_2$ in the Martian atmosphere, 4~orders
of magnitude less than proposed by
\citet{YungNG97},
would result in a pH acidic enough to attack solid rocks and precipitate
gypsum (CaSO$_4$*2H$_2$O), removing the SO$_2$ from the atmosphere.

Some temperature increase on early Mars is expected from geothermal heating
\citep{Squy93},
but by itself it is insufficient
\citep{Whit+95}.

With the above greenhouse and geothermal heating apparently incapable of
yielding liquid water on early Mars, given the illumination from a standard
solar model, let us consider the possibility of a non-standard, brighter
young Sun.  Figure~\ref{fig:solarflux} presents the relative flux values~$S$
throughout the Sun's past history
yielded by our mass-losing solar models with initial masses from $M_i = 1.01$
to~$1.07\;M_\odot$ (as well as that from a standard solar model).  The
requirement that the early Earth not lose its water via a moist greenhouse
effect leads to an upper limit of~$S \le 1.1$, corresponding to the upper
edge of the figures.  The lower luminosity constraints from the requirement
that liquid water be present 3.8~Gyr ago on early Mars
are shown by the vertical arrows; the heavy double arrow corresponds to
the limit $S \ge 0.86$ obtained by
\citet{Kast91},
and the lighter single arrow to his extreme (less probable)
limit $S \ge 0.80$.



Figure~\ref{fig:solarflux}a presents the solar flux~$S$ at Earth and Mars
(relative to their present flux) from the time of solar system formation
until the present,
for our ``exponential'' mass loss cases; the flux for the standard solar
model (without mass loss) is also shown, by the solid line.  Note that
the mass loss time
scale is between 0.755 and 0.551~Gyr, with initial mass loss
rates from $1.33 \times 10^{-11}$ to
$1.27 \times 10^{-10} \; M_\odot\,$yr$^{-1}$, for initial masses from 1.01
to $1.07\;M_\odot$, respectively.
The exponential decline as a function of time of these mass loss rates
means that they are generally consistent with the observations of
$\pi^{01}$~UMa, $\kappa^1$~Cet, and $\beta$~Com presented by
\citet{GaiGB00},
who obtained upper limits of $5 \times 10^{-11}$, $4 \times 10^{-11}$, 
and $4 \times 10^{-11} \; M_\odot\,$yr$^{-1}$, respectively,
for the mass loss rates of
these three young Sun-like stars (note that the $M_i = 1.07 \; M_\odot$ case
is only marginally consistent, lying very slightly above the
$\pi^{01}$~UMa limit: see Fig.~\ref{fig:mdot}).
Our results demonstrate that, for the ``exponential'' mass loss cases,
the $M_i = 1.07\;M_\odot$ case --- and only this case --- is
marginally consistent with the
\citet{Kast91}
Mars flux requirement $S \gtrsim 0.86$ at an age of
$\sim 3.8$~Gyr ago; if the Martian surface CO$_2$ pressure 3.8~Gyr ago
was either much lower or much higher than 5~bar, even the
$M_i = 1.07\;M_\odot$ case would be ruled out.  (If the
unlikely extreme Kasting flux limit of $S \gtrsim 0.80$ is used,
initial masses $1.03\;M_\odot \lesssim M_i \lesssim 1.07\;M_\odot$
would be permissible for a CO$_2$ pressure of 5 bars, and the
$M_i = 1.07\;M_\odot$ case would be marginally compatible with pressures
between $\sim 3$ and~$\sim 12$ bar.)

Since the semi-empirical mass loss law
$\dot M \propto t^{-2.00\pm0.52}$ recently presented by
\citet{Wood+02}
is strongly peaked at early times (as discussed in \S~\ref{sssec:mdotobs}),
it tends to yield a relatively small solar mass $\sim 3.8$~Gyr ago.  A
$t^{-2}$ mass loss relation, normalized by the present solar mass loss
rate, would yield $M(-3.8\;$Gyr$) \sim 1.001 \; M_\odot$; by tweaking
parameters to their extreme limits, one might obtain
$M(-3.8\;$Gyr$) \sim 1.02 \; M_\odot$.  Comparing the flux relations in
Figure~\ref{fig:solarflux} with the solar mass as a function of time in
Figure~\ref{fig:helio}, one can estimate that a solar mass of
$M(-3.8\;$Gyr$) \gtrsim 1.018 \; M_\odot$ is needed in order to satisfy the
\citet{Kast91}
Mars flux requirement $S(-3.8\;$Gyr$) \gtrsim 0.86$.
Thus the mass loss formula of
\citet{Wood+02}
might possibly be capable of satisfying the Mars flux requirement.
This possibility will be tested in a future paper
\citetext{A.~I.~Boothroyd \& \hbox{I.-J.}~Sackmann, in preparation}.

Figure~\ref{fig:solarflux}b
presents the extreme ``step-function'' mass loss
case.  This case has a constant mass loss rate for the first 1.6~Gyr
($\dot M = 5.69 \times 10^{-12}$ to 
$4.32 \times 10^{-11} \; M_\odot\,$yr$^{-1}$, for initial
solar masses of 1.01 to $1.07 \; M_\odot$, respectively),
and a low mass loss rate thereafter.  
These mass loss rates are all consistent with the stellar
mass loss observations of
\citet{GaiGB00}
for $\pi^{01}$~UMa, $\kappa^1$~Cet, and $\beta$~Com quoted above.
These ``step-function'' cases have a longer mass loss
timescale than the ``exponential'' one, and thus yield a higher solar
flux for the first 1.6~Gyr.  (Due to the way the ``step-function''
cases were defined, after the first 1.6~Gyr their solar flux is very close to
that of the standard solar model.)
Our results demonstrate that, for this extreme ``step-function'' mass loss
case, initial masses $1.04\;M_\odot \lesssim M_i \lesssim 1.07\;M_\odot$
are capable of yielding liquid water on Mars until 3.8~Gyr ago.

Figure~\ref{fig:solarflux}c similarly presents the radical ``linear'' mass loss
case.  This case has a high initial mass loss rate
($\dot M = 4.35 \times 10^{-12}$ to 
$3.04 \times 10^{-11} \; M_\odot\,$yr$^{-1}$, for initial
solar masses of 1.01 to $1.07 \; M_\odot$, respectively),
which remains relatively high throughout much of the Sun's lifetime
(since it declines linearly with time to reach the present solar mass
loss rate at the present time).
These mass loss cases are consistent with the stellar
mass loss observations of
\citet{GaiGB00}
for $\pi^{01}$~UMa, $\kappa^1$~Cet, and $\beta$~Com quoted above.
However, they are {\em not\/} consistent with the lunar rock
observations of the solar wind over the past 3~Gyr, violating this
latter constraint by an order of magnitude.
Our results demonstrate that, for this radical ``linear'' mass loss
case, initial masses $1.03\;M_\odot \lesssim M_i \lesssim 1.07\;M_\odot$
are capable of yielding liquid water on Mars until 3.8~Gyr ago.
The lower end of this range is mildly (but not significantly)
favored by the helioseismology; note that such cases with
$M_i \sim 1.04\;M_\odot$ have remarkably constant solar flux
over the first 3~Gyr.

\subsection{The Favored Cases of a Bright Young Sun} \label{ssec:favored}

A mass-losing solar model will always be brighter at birth than the
standard solar model, since the luminosity~$L_{ZAMS}$
at the zero age main sequence
(ZAMS) is roughly proportional to the mass to the fourth power
($L_{ZAMS} \propto M_i^4$).
For a mass-losing Sun, the orbital radii of
the planets varies inversely with the solar mass
($r_i \propto 1 / M_i$, due to conservation
of angular momentum); the initial flux at the planets is thus
proportional to the sixth power of the initial solar mass
($F_{ZAMS} \propto L_{ZAMS} / r_i^2 \propto M_i^6$).
Figure~\ref{fig:solarflux} illustrates the solar flux at the planets
as a function of time, demonstrating how much higher the early solar
flux at the planets is in the mass-losing cases than in the standard
(non-mass-losing) model.



\begin{figure}[t]
  \epsscale{0.49}
 \plotone{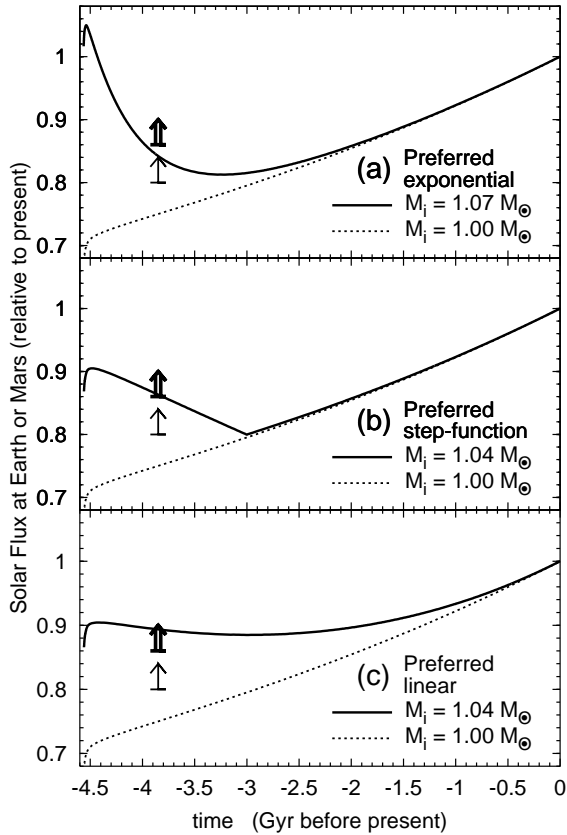}

\caption{Solar flux at the planets (relative to the present flux) as a
function of time for our preferred initial masses, for each type of mass
loss that we considered.  Heavy double arrows give the lower flux limit of
\citet{Kast91}
for the presence of water on early Mars; light single arrows give his
extreme lower flux limit (for a model with an unrealistically low
Martian surface albedo).}

 \label{fig:prefflux}

\end{figure}


For clarity, only the flux for
our ``preferred'' initial masses for each type of mass loss
are illustrated in Figure~\ref{fig:prefflux}.
Our preferred ``exponential'' case (with $M_i = 1.07 \; M_\odot$)
predicts a solar flux at the planets
about 5\% higher at birth than at present, considerably higher
than that indicated by the standard solar model (which predicts
a flux 29\% lower than at present).  At 3.8~Gyr ago, the flux for our
``exponential'' case would have been only 16\% lower than at present
(cf.\ 25\% for the standard model).
For our preferred ``step-function'' case (with $M_i = 1.04 \; M_\odot$),
the flux at the planets would have been only 10\% lower at birth
than at present (cf.\ 29\% for the standard model); at 3.8~Gyr
ago, the flux would have been only 14\% lower than at present
(cf.\ 25\% for the standard model).  For these ``exponential'' and
``step-function'' cases, the flux at the planets for the past 3~billion
years would be essentially the same as that predicted by the
standard solar model.
Our radical ``linear'' case (with $M_i = 1.04 \; M_\odot$)
would have had an almost constant solar flux at the planets
for the first 3~Gyr, namely, only 11\% lower than at present
(cf.\ 29\% to 12\% lower for the standard model); for this case,
the flux would be close to that predicted by the standard solar model
only during the last billion years.



\begin{figure}[t]
 \epsscale{0.67}
 \plotone{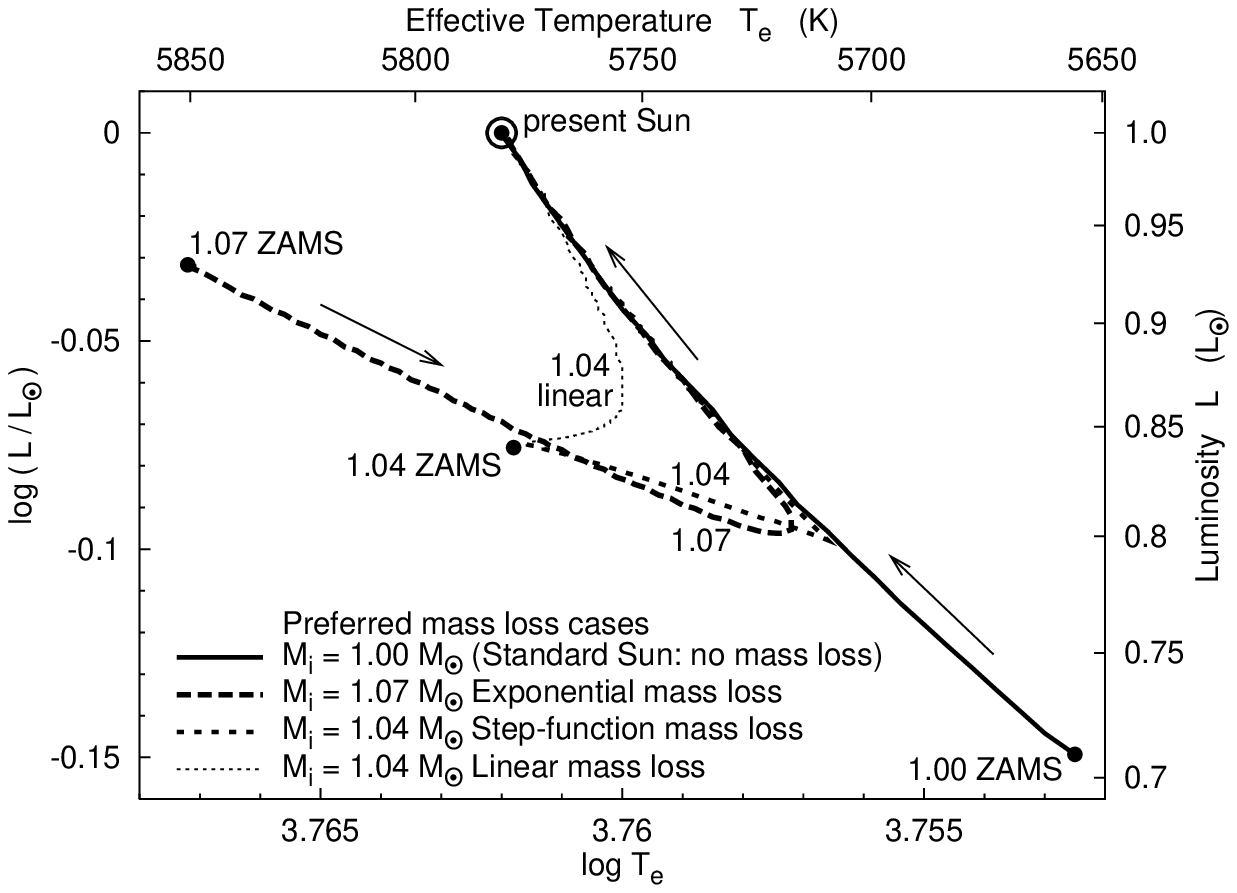}

\caption{Evolution in the H-R diagram of the standard solar model, and
of our preferred mass loss cases for each type of mass loss considered.
The ``ZAMS'' points shown are actually $\sim 50$~Myr subsequent to the
start of nuclear burning on the main sequence, i.e., the rapid loop
due to initial CN-cycle burning is omitted for clarity.}

 \label{fig:HRD}

\end{figure}


Figure~\ref{fig:HRD} presents the evolution in the HR~diagram of
our preferred exponential and step-function mass-losing cases
(heavy dashed and dot-dashed curves, respectively);
these cases are in agreement with the
helioseismic observations, with the existence of water on early Mars,
and with the lunar rock observations of solar mass loss.  (A radical linear
case, agreeing with the first two of these constraints
but disagreeing with the third one, is shown by the light dotted curve.)
For comparison, the standard solar model is also displayed (solid line).
Figure~\ref{fig:HRD} illustrates that the early evolution of mass-losing
solar
models is in the opposite direction in the HR~diagram to the standard
solar model: the mass-losing models initially
become less luminous and slightly redder (instead of more luminous and
slightly bluer).  Figure~\ref{fig:HRD} also illustrates that in the
past, the Sun's surface temperature changed only by negligible amounts
(1~or~2\%), both for the standard and the mass losing cases, in contrast
to the relatively large changes in the luminosity.

\section{Conclusions}   \label{sec:concl}

A slightly higher initial solar mass, producing a brighter young Sun,
turns out to be a viable explanation for warm temperatures
on early Earth and Mars, that otherwise are difficult to account for
(particularly for Mars).  Such a higher initial solar mass leaves a
fingerprint on the Sun's present internal structure that is
large enough to be detectable in principle via helioseismic observations. 
Our computations demonstrated that all 21 of the mass-losing solar
models that we considered were consistent with
the helioseismic observations; in fact, our preferred mass-losing cases
were in marginally (though not significantly) better agreement with the
helioseismology than the standard solar model was.
However, there are still significant uncertainties in the
observed solar composition and in the input physics on which the solar
models are based; these uncertainties have a slightly larger effect on the
Sun's present internal structure than the fingerprint left from
early solar mass loss.  Future improvements by a factor of~2 or so
in the accuracy of
these input parameters could reduce the size of the uncertainties
below the level of the fingerprints left by a more massive, brighter young
Sun, allowing one to determine whether early solar mass loss took place
or not.  Also urgently needed are more measurements of mass loss rates from
other young stars similar to the young Sun, and more measurements from
our solar system that can be used to estimate the solar wind
in the past.

\acknowledgements

We are indebted to Prof.\ Marc H. Pinsonneault for helpful discussions on
diffusion and for providing us with his diffusion code.  We are grateful to
Prof.\ Sarbani Basu for discussions of helioseismology, and for providing
us with the current helioseismic reference model; we are also grateful to
Prof. Dimitri M. Mihalas, for providing us with his equation-of-state code.
We wish to thank Prof.\ Charles A. Barnes, Prof.\ Yuk L. Yung, and
Dr.\ Mimi F. Gerstell for thoughtful discussions and encouragement.
We wish to acknowledge the support provided by Prof.\ Thomas A. Tombrello,
Chairman of the Division of Physics, Math, and Astronomy,
and Prof.\ Robert D. McKeown, Head of the W.~K.~Kellogg Radiation Laboratory.
One of us (I.-J.~S.) wishes to thank Alexandra R. Christy, her daughter,
and Prof.\ Robert F. Christy, her husband, for their
supportiveness, and Robert F. Christy
for critical analysis and helpful comments.  One of us (A.~I.~B.) wishes to
thank Prof.\ Peter G. Martin and Prof.\ J. Richard Bond for their support,
and M.~Elaine Boothroyd, his wife, for her patience and encouragement.
This work was supported by a grant \hbox{NAG5-7166} from the Sun-Earth
Connection Program of the Supporting Research and Technology
and Suborbital Program in Solar Physics of the National Aeronautics and
Space Administration, and by the National Science Foundation
grant \hbox{NSF-0071856} to the Kellogg Radiation Laboratory.

\clearpage

\begin{deluxetable}{lccccccccc@{\hspace{0pt}}c@{\hspace{0pt}}cccccc}

\tabletypesize{\footnotesize}

\rotate

\tablewidth{0pt}

\tablecaption{Characteristics of Our Solar
 Models\tablenotemark{a}\label{tab:results}}

\tablehead{
\colhead{} &
 \colhead{} & \colhead{} & \colhead{} & \colhead{} & 
 \colhead{$R_{ce}$} &
 \multicolumn{2}{c}{rms $\delta c/c$ for:} &
 \colhead{rms} &
 \multicolumn{3}{c}{relative rms} &
 \colhead{} & \colhead{} & \colhead{} & \colhead{} & \colhead{} \\
\cline{7-8} \cline{10-12}
\colhead{Solar Model} &
 \colhead{$\alpha$} & \colhead{$Z_0$} & \colhead{$Y_0$} & \colhead{$Y_e$} &
 \colhead{($R_\odot$)} &
 \colhead{all-$r$} & \colhead{$< 0.6$} &
 \colhead{$\delta\rho/\rho$} &
 \colhead{vs.} & \colhead{$\Delta c/c$} & \colhead{$\Delta\rho/\rho$} &
 \colhead{$f_{\rm Li}$} & \colhead{$f_{\rm Be}$} & \colhead{$\Phi_{\rm Cl}$} &
 \colhead{$\Phi_{\rm Ga}$} & \colhead{$\Phi_{\rm B}$}
}

\startdata
%
1. Fine-zoned Reference\tablenotemark{b} &
 1.817 & .02030 & .2760 & .2424 & .7135 & .00133 & .00085 & .01698 &
  & \nodata & \nodata & 24.24 & 1.008 & 7.87 & 133.7 & 5.31 \\
\tableline
2. Fine-zoned $1.01\;M_\odot$ exp &
 1.818 & .02025 & .2754 & .2425 & .7133 & .00116 & .00074 & .01533 &
 \phn1 & .00028 & .00179 & 19.22 & 1.009 & 7.87 & 133.7 & 5.32 \\
%
3. Fine-zoned $1.02\;M_\odot$ exp &
 1.819 & .02017 & .2748 & .2425 & .7133 & .00126 & .00074 & .01544 &
 \phn1 & .00017 & .00158 & 17.43 & 1.009 & 7.87 & 133.7 & 5.32 \\
%
4. Fine-zoned $1.03\;M_\odot$ exp &
 1.819 & .02012 & .2743 & .2426 & .7133 & .00118 & .00070 & .01437 &
 \phn1 & .00034 & .00276 & 17.81 & 1.009 & 7.88 & 133.7 & 5.32 \\
%
5. Fine-zoned $1.04\;M_\odot$ exp &
 1.819 & .02007 & .2738 & .2427 & .7132 & .00118 & .00068 & .01398 &
 \phn1 & .00034 & .00311 & 19.91 & 1.012 & 7.89 & 133.8 & 5.33 \\
%
6. Fine-zoned $1.05\;M_\odot$ exp &
 1.821 & .02003 & .2734 & .2428 & .7131 & .00112 & .00065 & .01340 &
 \phn1 & .00039 & .00366 & 22.96 & 1.026 & 7.91 & 133.9 & 5.35 \\
%
7. Fine-zoned $1.06\;M_\odot$ exp &
 1.821 & .01999 & .2730 & .2428 & .7130 & .00105 & .00063 & .01247 &
 \phn1 & .00049 & .00464 & 26.99 & 1.069 & 7.92 & 133.9 & 5.35 \\
%
8. Fine-zoned $1.07\;M_\odot$ exp &
 1.823 & .01997 & .2726 & .2429 & .7128 & .00096 & .00062 & .01162 &
 \phn1 & .00060 & .00550 & 31.82 & 1.173 & 7.94 & 134.1 & 5.37 \\
\tableline
9. Coarse-zoned Reference\tablenotemark{b,c} &
 1.814 & .02030 & .2760 & .2424 & .7136 & .00140 & .00091 & .01772 &
 \phn1\tablenotemark{c} & .00008 & .00074 &
 24.36 & 1.010 & 7.89 & 133.8 & 5.33 \\
\tableline
10. $1.01\;M_\odot$ exp &
 1.816 & .02024 & .2754 & .2427 & .7134 & .00128 & .00081 & .01641 &
 \phn9 & .00016 & .00134 & 19.22 & 1.009 & 7.91 & 134.0 & 5.35 \\
%
11. $1.02\;M_\odot$ exp &
 1.816 & .02016 & .2748 & .2429 & .7134 & .00124 & .00077 & .01569 &
 \phn9 & .00024 & .00208 & 17.26 & 1.009 & 7.89 & 133.8 & 5.33 \\
%
12. $1.03\;M_\odot$ exp &
 1.816 & .02010 & .2742 & .2431 & .7134 & .00120 & .00074 & .01515 &
 \phn9 & .00030 & .00263 & 17.82 & 1.009 & 7.89 & 133.8 & 5.33 \\
%
13. $1.04\;M_\odot$ exp &
 1.816 & .02005 & .2738 & .2433 & .7133 & .00115 & .00070 & .01448 &
 \phn9 & .00037 & .00332 & 19.72 & 1.012 & 7.90 & 133.8 & 5.34 \\
%
14. $1.05\;M_\odot$ exp &
 1.817 & .02001 & .2734 & .2433 & .7133 & .00111 & .00067 & .01379 &
 \phn9 & .00045 & .00402 & 22.82 & 1.026 & 7.91 & 133.9 & 5.35 \\
%
15. $1.06\;M_\odot$ exp &
 1.818 & .01997 & .2730 & .2434 & .7131 & .00105 & .00065 & .01300 &
 \phn9 & .00053 & .00484 & 27.47 & 1.068 & 7.93 & 134.0 & 5.36 \\
%
16. $1.07\;M_\odot$ exp &
 1.819 & .01995 & .2726 & .2434 & .7131 & .00101 & .00063 & .01217 &
 \phn9 & .00062 & .00569 & 34.25 & 1.170 & 7.95 & 134.1 & 5.38 \\
\tableline
17. $1.01\;M_\odot$ step &
 1.816 & .02019 & .2751 & .2426 & .7134 & .00128 & .00080 & .01626 &
 \phn9 & .00018 & .00149 & 19.57 & 1.009 & 7.87 & 133.7 & 5.32 \\
%
18. $1.02\;M_\odot$ step &
 1.815 & .02007 & .2742 & .2432 & .7135 & .00124 & .00077 & .01572 &
 \phn9 & .00027 & .00204 & 18.94 & 1.009 & 7.86 & 133.6 & 5.31 \\
%
19. $1.03\;M_\odot$ step &
 1.815 & .01999 & .2734 & .2435 & .7135 & .00117 & .00072 & .01463 &
 \phn9 & .00035 & .00316 & 20.87 & 1.011 & 7.88 & 133.7 & 5.32 \\
%
20. $1.04\;M_\odot$ step &
 1.816 & .01993 & .2728 & .2437 & .7133 & .00109 & .00067 & .01347 &
 \phn9 & .00048 & .00435 & 24.36 & 1.025 & 7.90 & 133.8 & 5.34 \\
%
21. $1.05\;M_\odot$ step &
 1.818 & .01988 & .2722 & .2437 & .7131 & .00101 & .00064 & .01207 &
 \phn9 & .00063 & .00580 & 29.75 & 1.074 & 7.94 & 134.0 & 5.37 \\
%
22. $1.06\;M_\odot$ step &
 1.821 & .01985 & .2716 & .2437 & .7129 & .00095 & .00064 & .01061 &
 \phn9 & .00079 & .00733 & 37.76 & 1.213 & 7.98 & 134.3 & 5.40 \\
%
23. $1.07\;M_\odot$ step &
 1.823 & .01982 & .2711 & .2436 & .7126 & .00089 & .00068 & .00916 &
 \phn9 & .00095 & .00883 & 49.63 & 1.534 & 8.01 & 134.4 & 5.43 \\
\tableline
24. $1.01\;M_\odot$ linear &
 1.818 & .02021 & .2749 & .2427 & .7132 & .00116 & .00070 & .01486 &
 \phn9 & .00035 & .00293 & 19.70 & 1.009 & 7.94 & 134.1 & 5.37 \\
%
25. $1.02\;M_\odot$ linear &
 1.818 & .01997 & .2731 & .2429 & .7131 & .00122 & .00069 & .01427 &
 \phn9 & .00046 & .00356 & 18.49 & 1.009 & 7.92 & 134.0 & 5.36 \\
%
26. $1.03\;M_\odot$ linear &
 1.816 & .01972 & .2713 & .2439 & .7133 & .00115 & .00069 & .01293 &
 \phn9 & .00060 & .00496 & 20.11 & 1.011 & 7.87 & 133.6 & 5.32 \\
%
27. $1.04\;M_\odot$ linear &
 1.817 & .01959 & .2700 & .2445 & .7133 & .00105 & .00070 & .01084 &
 \phn9 & .00082 & .00713 & 23.34 & 1.027 & 7.90 & 133.8 & 5.35 \\
%
28. $1.05\;M_\odot$ linear &
 1.817 & .01947 & .2687 & .2450 & .7132 & .00099 & .00078 & .00858 &
 \phn9 & .00106 & .00958 & 28.35 & 1.081 & 7.95 & 134.1 & 5.38 \\
%
29. $1.06\;M_\odot$ linear &
 1.821 & .01937 & .2675 & .2452 & .7128 & .00099 & .00094 & .00640 &
 \phn9 & .00132 & .01200 & 36.71 & 1.250 & 8.00 & 134.4 & 5.43 \\
%
30. $1.07\;M_\odot$ linear &
 1.824 & .01929 & .2663 & .2452 & .7125 & .00108 & .00114 & .00432 &
 \phn9 & .00161 & .01476 & 48.90 & 1.631 & 8.07 & 134.7 & 5.48 \\
\tableline
31. GS98 reference\tablenotemark{d} &
 1.775 & .01912 & .2759 & .2418 & .7155 & .00188 & .00120 & .02121 &
 \phn9\tablenotemark{d} & .00060 & .00378 &
 18.37 & 1.009 & 7.73 & 132.6 & 5.24 \\
\tableline
32. GS98 $1.01\;M_\odot$ exp &
 1.776 & .01906 & .2752 & .2421 & .7154 & .00178 & .00111 & .02014 &
 31 & .00035 & .00123 & 14.85 & 1.008 & 7.78 & 132.9 & 5.28 \\
%
33. GS98 $1.02\;M_\odot$ exp &
 1.776 & .01899 & .2746 & .2422 & .7154 & .00175 & .00105 & .01929 &
 31 & .00035 & .00201 & 13.70 & 1.008 & 7.72 & 132.5 & 5.23 \\
%
34. GS98 $1.03\;M_\odot$ exp &
 1.776 & .01893 & .2741 & .2425 & .7153 & .00168 & .00101 & .01864 &
 31 & .00044 & .00267 & 14.47 & 1.008 & 7.73 & 132.6 & 5.24 \\
%
35. GS98 $1.04\;M_\odot$ exp &
 1.777 & .01889 & .2736 & .2427 & .7153 & .00162 & .00097 & .01799 &
 31 & .00049 & .00332 & 16.46 & 1.010 & 7.74 & 132.6 & 5.25 \\
%
36. GS98 $1.05\;M_\odot$ exp &
 1.778 & .01885 & .2732 & .2428 & .7152 & .00157 & .00092 & .01725 &
 31 & .00056 & .00407 & 19.58 & 1.025 & 7.75 & 132.7 & 5.26 \\
%
37. GS98 $1.06\;M_\odot$ exp &
 1.778 & .01881 & .2728 & .2428 & .7151 & .00152 & .00088 & .01652 &
 31 & .00063 & .00481 & 24.15 & 1.062 & 7.77 & 132.7 & 5.27 \\
%
38. GS98 $1.07\;M_\odot$ exp &
 1.780 & .01879 & .2724 & .2428 & .7150 & .00145 & .00084 & .01573 &
 31 & .00070 & .00561 & 30.84 & 1.160 & 7.78 & 132.8 & 5.28 \\
\tableline
Worst variant model\tablenotemark{e} &
 \nodata & \nodata & \nodata & \nodata & \nodata & .00366 & .00267 & .03376 &
 \nodata & .00184 & .02614 &
 \nodata & \nodata & \nodata & \nodata & \nodata \\
%
\enddata

\tablenotetext{a}{ Mixing length parameter~$\alpha$, pre-solar
 metallicity~$Z_0$ and helium mass fraction~$Y_0$, present envelope
 helium abundance~$Y_e$, position~$R_{ce}$ of the base of envelope
 convection, rms fractional sound speed and density differences relative
 to the Sun's inferred helioseismic profiles and relative to the reference
 standard solar model, pre-main-sequence lithium depletion
 factor~$f_{\rm Li}$ and beryllium depletion factor~$f_{\rm Be}$,
 predicted capture rates (in SNU) $\Phi_{\rm Cl}$ and
 $\Phi_{\rm Ga}$ for chlorine and gallium experiments, respectively, and
 predicted flux~$\Phi_{\rm B}$ of \iso8B neutrinos (in units of
 $10^6$~cm~s$^{-1}$).  Seven different cases with initial masses from
 1.01 to $1.07\;M_\odot$ are shown for each of the three mass loss types,
 namely, exponential (``exp''), step-function (``step''), and
 linear (``linear'').}

\tablenotetext{b}{ Reference standard solar model: OPAL EOS at
 $\log\,\rho \gtrsim -1.5$, MHD EOS at $\log\,\rho \lesssim -2$,
 high-T opacities $\kappa_{\rm OPAL:GN93}$ interpolated in
 $Z_\kappa = Z_h \equiv Z_0 \, [ \sum_{heavy} X_i ] / [ \sum_{heavy} (X_i)_0 ]$
 as well as in ``excess'' C and~O (such that
 ${\rm C}_{ex} + {\rm O}_{ex} \equiv {\rm CO}_{ex} = Z - Z_\kappa$),
 low-T opacities $\kappa_{\rm Alexander}$, NACRE nuclear rates,
 gravitational settling of He and heavy elements, $Z/X = 0.0245$,
 $L_\odot = 3.854 \times 10^{33}$~erg~s$^{-1}$,
 $R_\odot = 695.98$~Mm, and $t_\odot = 4.6$~Gyr; both fine-zoned and
 coarse-zoned cases were computed.}

\tablenotetext{c}{ This reference standard solar model and all subsequent
 models in the table use the coarse zoning.  Note that the relative rms
 values for this coarse-zoned reference standard solar model compare it
 to the fine-zoned reference model.}

\tablenotetext{d}{ This ``GS98'' alternate standard solar model (and the
 following ``GS98'' mass-losing models in the table) has
 $Z/X = 0.023$ with the appropriate OPAL opacities $\kappa_{\rm OPAL:GS98}$,
 using the OPAL EOS at $\log\,T \gtrsim 4.0$ and the MHD EOS at
 $\log\,T \lesssim 3.9$; otherwise it is the same as the coarse-zoned
 reference standard solar model.  The relative rms values for
 this ``GS98'' alternate standard solar model compare it to the
 coarse-zoned reference model.}

\tablenotetext{e}{ Worst rms and relative rms values from standard solar
 models with ``reasonable'' variations of the solar input parameters, from
 \citet{BS02}.}

\end{deluxetable}

\end{document}